\documentclass[reprint,pra]{revtex4-1}

\usepackage[english]{babel}
\usepackage[T1]{fontenc}
\usepackage[utf8]{inputenc}
\usepackage{amsmath}
\usepackage{mathtools}
\usepackage{color}
\usepackage{graphicx}

\newcommand{\iPsi}{\mathit{\Psi}}

\newcommand{\pder}[3]{\frac{\partial^{#3} #1}{\partial #2^{#3}}}

\newcommand{\mpder}[6]{\frac{\partial^{#6} #1}{\partial #2^{#3} \partial #4^{#5}}}

\newcommand{\at}[1]{\big|_{#1}}
\newcommand{\mat}[1]{\Big|_{#1}}
\newcommand{\bat}[1]{\bigg|_{#1}}

\newcommand{\mean}[1]{\langle#1\rangle}
\newcommand{\meanb}[1]{\left\langle#1\right\rangle}

\begin{document}

\title{Corrected perturbation theory for transverse-electric whispering-gallery modes in deformed microdisks}

\author{Manuel Badel}
\altaffiliation{manuel.badel@st.ovgu.de}
\author{Jan Wiersig}
\altaffiliation{jan.wiersig@ovgu.de}
\affiliation{Institut für Physik, Otto-von-Guericke-Universität Magdeburg, Postfach 4120, D-39016 Magdeburg, Germany}
\date{\today}

\begin{abstract}
The perturbation theory by L. Ge \textit{et al.} [Phys. Rev. A 87, 023833] for transverse-electric polarized modes in weakly deformed microdisks omits terms related to the variation of the normal derivative of the magnetic field along the boundary. Here, we show that these terms are necessary to accurately describe microdisks with a strongly winding boundary. In particular, it is demonstrated that the corrected perturbation theory allows to describe the counterintuitive phenomenon of $Q$-factor enhancement due to weak boundary deformation. We discuss in detail the microflower cavity and the lima\c{c}on cavity. Good agreement of the corrected perturbation theory with full numerical results is observed. 
\end{abstract}

\pacs{Valid PACS appear here}
\maketitle

\section{Introduction}
\label{Introduction}
Optical microcavities (or microresonators) have attracted increasing attention because of the possibility to resonantly confine light on small spatial scales for a long time~\cite{Vahala03}. 
There is a wealth of applications such as tiny sensors for the detection of specific gases, gas compositions, and even single nanoparticles~\cite{ZOX10}. Furthermore, microcavity sensors are versatile and highly sensible to different physical quantities, like electromagnetic fields, temperatures, pressures, and forces as reviewed in Refs.~\cite{VY12,FSV15}.
Lasing is another important application that can be accomplished by different cavity designs. Two prominent ones are micropillars~\cite{Reitzenstein06,Ulrich06} that emit light mainly along their symmetry axis and microdisks~\cite{MLSGPL92,SLMMOL93,WRS11} that emit mainly in the disk plane. 

Microdisks belong to the class of whispering-gallery cavities which are based on the successive total internal reflection of light at the curved boundary of the cavity. The rotational symmetry ensures that the corresponding optical modes, the whispering-gallery modes (WGMs), have very low losses which is usually quantified by the quality factor ($Q$-factor). However, the rotational symmetry also implies isotropic light emission in the plane of the disk which is especially disadvantageous for lasing. 

One possibility to solve this problem is by deforming the boundary as proposed in Ref.~\cite{ND97} and experimentally confirmed in Ref.~\cite{GCNNSFSC98}. Emission into a single direction with small angular spread has been achieved, e.g., with the lima\c{c}on cavity~\cite{WH08,SHH09,WYD09,SCL09,YWD09,AHE12}, the notched cavity~\cite{BBS06b,WYY10}, and the short-egg geometry~\cite{SBS15}.
Boundary deformation also helps to achieve efficient broadband transmission from waveguide to whispering-gallery microcavity and back~\cite{JSZ17}.
Unfortunately, deformation often leads to the phenomenon of $Q$-spoiling, i.e., to an undesired decrease of the $Q$-factor~\cite{Noeckel94}.
However, $Q$-spoiling can also be exploited for mode selection in high-power single-frequency lasers by damaging unwanted modes~\cite{FB01,FB02,BBS06b,SAS13}. 
Various other aspects of deformed microdisk cavities, such as non-Hermitian effects and chaotic ray dynamics, are reviewed in Ref.~\cite{CW15}.

In microdisks two types of polarization can be distinguished, the transverse-magnetic (TM) polarization where the electric field is perpendicular to the disk plane, and the transverse-electric (TE) polarization where this is the case for the magnetic field.
A perturbation theory (PT) for TM-polarized WGMs in weakly deformed microdisks, which possess at least one mirror-reflection symmetry, was proposed by Dubertrand \textit{et al.}~\cite{DBD08}. The general applicability of the theory, in some cases with additional generalizations or modifications, was tested multiple times~\cite{Wiersig12,KW14,Badel15,KraftW16,KW16c,WK17,Badel17,KYW18}. There, it was found that the theory works well, and for some deformations better than expected, but characteristic differences between modes from different symmetry classes could not be reproduced~\cite{Badel15,Badel17}.
The analogous theory for TE-polarized WGMs had been derived by Ge \textit{et al.}~\cite{GSR13} to explain drastic output sensitivities to very small boundary deformations. 

In the present paper, we test the applicability of the TE PT by studying various boundary deformations with special attention to the so-called microflower cavity. This cavity was numerically studied by Boriskina \textit{et al.}~\cite{BBS03,BBS06b} because of its beneficial mode selection properties. Moreover, for small boundary deformation it can exhibit a peculiar $Q$-factor enhancement, which is the opposite of $Q$-spoiling. Our study reveals that the PT in Ref.~\cite{GSR13} is not able to reproduce this observation, and that it is generally less accurate than its TM counterpart. We explain why this is the case and derive a correction of the TE theory which shines light onto the phenomenon of $Q$-factor enhancement.

The paper is organized as follows. Sections~\ref{S_CircularMicrodisk} and \ref{S_PerturbationTheories} briefly review the mathematics and mode properties of circular microdisks as well as the basis of the perturbation theories. The corrections for TE polarization are discussed in Sec.~\ref{S_CorrectionOfTheTETheory}; a detailed derivation can be found in App.~\ref{A_DerivationOfTheCorrections}. 
The regime of applicability of the perturbation series is determined in Sec.~\ref{sec:applicabilit}.
Section~\ref{S_ResultsForTheCorrectedTheory} shows the results of the corrected PT compared to the uncorrected one and full numerical calculations, first for the microflower and then the lima\c{c}on cavity. A conclusion is given in Sec.~\ref{S_Conclusion}.

\section{Circular microdisk}
\label{S_CircularMicrodisk}
Microdisks are two-dimensional systems because their height is negligible in comparison to their other elongations. As explained in Refs.~\cite{SNB05,LDA07,BDM09} the $z$-direction (perpendicular to the disk plane) can be separated out using the effective-index approximation which goes hand in hand with a decoupling of the two types of polarization. 

The optical modes are the solutions of Maxwell's equations with harmonic time dependence $\exp{(-i\omega t)}$ where $\omega$ is the frequency. With the effective refractive index $n_j$, the two-dimensional mode equation in polar coordinates is 
	\begin{equation}
		\Big( \Delta_{r, \varphi} +n_j^2 k^2 \Big) \iPsi_j (r, \varphi) = 0.
	\label{E_ModeEquation}
	\end{equation}
$\iPsi$ is the electric field $E_z$ for TM polarization or the magnetic field $B_z$ for TE polarization, the other field components are of no further interest because they can be calculated from the $z$-component of the fields. The index $j = 1, 2$ denotes the inside or outside of the microdisk, respectively, $k = \omega/c$ is the wave number of the mode, and $c$ is the speed of light in vacuum.

The differential equation has to be accompanied by three boundary conditions. The first one requires the continuity of the wave function across the boundary $r_b (\varphi)$,
	\begin{equation}
		\iPsi_1 \at{r_b} = \iPsi_2 \at{r_b}.
	\label{E_BoundaryCondition_One}
	\end{equation}
The second boundary condition describes the behavior of the derivative of the wave functions across the boundary, which differ for both types of polarization. While the TM polarization's condition accounts for continuity, the TE polarization's condition demands \textit{dis}continuity,
	\begin{subequations}
		\begin{align}
			\pder{\iPsi_1}{\vec{\nu}}{} \bat{r_b} & = \pder{\iPsi_2}{\vec{\nu}}{} \bat{r_b}, \; \text{for TM};
		\label{E_BoundaryCondition_Two_TM} \\
			\frac{1}{n_1^2} \pder{\iPsi_1}{\vec{\nu}}{} \bat{r_b} & = \frac{1}{n_2^2} \pder{\iPsi_2}{\vec{\nu}}{} \bat{r_b}, \; \text{for TE},
		\label{E_BoundaryCondition_Two_TE}
		\end{align}
	\end{subequations}
where $\vec{\nu} (\varphi)$ is the normal vector to the boundary. Finally, the last boundary condition is the outgoing-wave condition far away from the cavity,
	\begin{equation}
		\iPsi_2 (r, \varphi) = \frac{\exp(i n_2 k r)}{\sqrt{n_2 k r}} g (\varphi),
	\label{E_OutgoingWaveCondition}
	\end{equation}
with an angular distribution $g (\varphi)$.

Only special cavity shapes can be treated analytically, like the circular cavity with the radius $r_b = R$. For simplicity let $n_1 = n$ and $n_2 = 1$, i.e., the cavity is surrounded by air. Because the substitutions $n_2 k = \tilde{k}$ and $n_1/n_2 = n$ always provide this simplification, no generality is lost. The modes in the circular cavity are typically twofold degenerate but this degeneracy can be removed by utilizing symmetric and antisymmetric wave functions. The symmetric inner and outer wave functions are
	\begin{equation}
		\begin{aligned}
			\iPsi_1 (r, \varphi) & = \tilde{a}_m J_m (n k r) \cos(m \varphi), \; \text{for} \; r \le R, \\
			\iPsi_2 (r, \varphi) & = \tilde{b}_m H_m^{(1)} (k r) \cos(m \varphi), \; \text{for} \; r > R,
		\end{aligned}
	\label{E_CircleWaveFunctions}
	\end{equation}
respectively. $J_m$ and $H_m^{(1)}$ are the Bessel and Hankel functions of the first kind and integer order $m \geq 0$; the indicator $^{(1)}$ for the first kind of the Hankel function will be omitted from now on. The antisymmetric wave functions are comprised of sines instead of cosines. Note that the wave functions~\eqref{E_CircleWaveFunctions} automatically fulfill the outgoing-wave condition~\eqref{E_OutgoingWaveCondition} due to the Hankel function's asymptotic behavior for large arguments. Below, the symmetric solutions are also called even-parity modes and the antisymmetric ones odd-parity modes.

With the wave functions~\eqref{E_CircleWaveFunctions} one can extract the conditional equations, which are identical for both parities, 
	\begin{subequations}
		\begin{align}
			S_m (k R) & = 0, \; \text{for TM} \ , \\
			T_m (k R) & = 0, \; \text{for TE} 
		\label{E_ConditionalEquation_TE}
		\end{align}
	\label{E_ConditionalEquations}
	\end{subequations}
with the definitions
		\begin{eqnarray}
                  \label{eq:Sm}
			S_m (k R) & = & n \frac{\dot{J}_m}{J_m} \big( n k R \big) -\frac{\dot{H}_m}{H_m} \big( k R \big) \ , \\
                  \label{eq:Tm}
			T_m (k R) & = & \frac{1}{n} \frac{\dot{J}_m}{J_m} \big( n k R \big) -\frac{\dot{H}_m}{H_m} \big( k R \big)
		\end{eqnarray}
by inserting them into the boundary conditions~\eqref{E_BoundaryCondition_One} and \eqref{E_BoundaryCondition_Two_TM} or \eqref{E_BoundaryCondition_Two_TE}, for $r_b = R$ respectively, and then dividing the latter by the first. Note that the dot defines the derivative with respect to the argument of the function, i.e., $nkr$ for $J_m$, $kr$ for $H_m$, and later $\varphi$ for $r_b$. Hence, the chain rule has already been applied in all formulas.

The conditional equations~\eqref{E_ConditionalEquations} provide complex wave numbers $k_{l, m} = k_r +i k_i$ and frequencies $\omega_{l, m} = ck_{l, m}$ which can be labeled by the mode numbers $l$ and $m$. $l$ is the radial mode number that counts the intensity maxima along the radial direction while $m$, the azimuthal mode number, does the same for half the maxima in angular direction; one period of the wave function in $\varphi$ produces two maxima for the intensity. The real part of the complex frequency determines the usual frequency while the imaginary part, which has to be negative, determines the lifetime. If the imaginary part is close to zero, as it usually is for WGMs, the corresponding mode is quasi-stationary.

\section{Perturbation theories}
\label{S_PerturbationTheories}
The boundary of the deformed microdisk is assumed to be of the form
	\begin{equation}
		r_b (\varphi) = R \Big[ 1 +\varepsilon f (\varphi) \Big]
	\label{E_Deformation}
	\end{equation}
with the dimensionless deformation strength $\varepsilon$ and the deformation function $f (\varphi)$. In general the deformation function is arbitrary, a TM theory for a fully asymmetric cavity can be found in Ref.~\cite{KW16c}.

For the perturbation theories by Dubertrand \textit{et al.}~\cite{DBD08} and Ge \textit{et al.}~\cite{GSR13}, however, symmetric deformation functions with $f (-\varphi) = f (\varphi)$ are considered to circumvent degeneracies. Then the symmetric wave functions
	\begin{widetext}
	\begin{equation}
		\begin{aligned}
			\iPsi_1 (r, \varphi) & = \frac{J_m (n k r)}{J_m (n k R)} \cos(m \varphi) +\sum_{p \neq m}^{} a_p \frac{J_p (n k r)}{J_p (n k R)} \cos(p \varphi), \; \text{for} \; r \le r_b (\varphi), \\
			\iPsi_2 (r, \varphi) & = \Big( 1 +b_m \Big) \frac{H_m (k r)}{H_m (k R)} \cos(m \varphi) +\sum_{p \neq m} \Big( a_p +b_p \Big) \frac{H_p (kr)}{H_p (k R)} \cos(p \varphi), \; \text{for} \; r > r_b (\varphi)
		\end{aligned}
	\label{E_DeformationWaveFunctions}
	\end{equation}
	\end{widetext}
serve as an ansatz, which are expansions in the symmetric wave functions of the circular disk. Antisymmetric wave functions can be used as well by substituting the cosines with sines. The complex frequencies and coefficients have to be series in $\varepsilon$, here up to the second order,
	\begin{equation}
		x = x_0 +\varepsilon x_1 +\varepsilon^2 x_2 +O \big( \varepsilon^3 \big),
	\label{E_EigenvalueExpansion}
	\end{equation}
	\begin{equation}
		a_p = \varepsilon a_{1, p} +\varepsilon^2 a_{2, p} +O \big( \varepsilon^3 \big),
	\label{E_CoefficientAExpansion}
	\end{equation}
	\begin{equation}
		b_p = \varepsilon b_{1, p} +\varepsilon^2 b_{2, p} +O \big( \varepsilon^3 \big).
	\label{E_CoefficientBExpansion}
	\end{equation}
The dimensionless frequency $x = \omega R/c = k R$ is used for convenience.
Note that this expansion is very different from alternative approaches where only the mode equation is expanded in first order in the frequency; see, e.g., for microspheres in Ref.~\cite{LLY90}.

The frequencies and coefficients can be calculated by inserting the ansatz~\eqref{E_DeformationWaveFunctions} and expansions~\eqref{E_EigenvalueExpansion}-\eqref{E_CoefficientBExpansion} into the boundary conditions~\eqref{E_BoundaryCondition_One} and \eqref{E_BoundaryCondition_Two_TM} or \eqref{E_BoundaryCondition_Two_TE}, expanding everything at $\varepsilon = 0$, then sorting the equations by power and finally analyzing those with Fourier harmonics. For a more detailed explanation that may include some information left out by Dubertrand \textit{et al.} in Ref.~\cite{DBD08} and Ge \textit{et al.} in Ref.~\cite{GSR13} we refer the reader to App.~\ref{A_DerivationOfTheCorrections}.

It is mentioned that the ansatz~\eqref{E_DeformationWaveFunctions} relies on the applicability of the Rayleigh hypothesis. In our case, the hypothesis states that the cavity's outer solutions only consist of Hankel functions of the first kind. A related problem has been studied by van den Berg and Fokkema in Ref.~\cite{BergFokkema79} in regard to scattering by a two-dimensional obstacle with zero Dirichlet boundary conditions. As the scattering problem is connected to the mode structure of a microcavity~\cite{Landau08,TSSJ05} we can use later the results from van den Berg and Fokkema to roughly estimate the reliability of our results.

\section{Correction of the TE perturbation theory}
\label{S_CorrectionOfTheTETheory}
For the purpose of this paper, the important part in Ref.~\cite{GSR13} is that for the TE polarization's second boundary condition
	\begin{equation}
		\frac{1}{n^2} \pder{\iPsi_1}{r}{} \bat{r_b} -\pder{\iPsi_2}{r}{} \bat{r_b} = 0
	\label{E_BoundaryCondition_Two_TE_Ge}
	\end{equation}
	is used, meaning that the normal derivative in condition~\eqref{E_BoundaryCondition_Two_TE} was interchanged with a radial one. In the TM PT this was done by Dubertrand \textit{et al.}~\cite{DBD08} for the reason that parallel \textit{and} normal derivatives are continuous along the boundary, see Eqs.~\eqref{E_BoundaryCondition_One} and \eqref{E_BoundaryCondition_Two_TM}, which ensures that the direction of the derivative can be chosen freely. But the second boundary condition~\eqref{E_BoundaryCondition_Two_TE} for the TE polarization is \textit{not continuous}, which means that the above simplification cannot be used.

Instead, the normal derivative must be accounted for, which, as a directional derivative, can be written as the scalar product of the direction and the gradient. In polar coordinates the product is
	\begin{equation}
		\begin{split}
			\pder{\iPsi_j}{\vec{\nu}}{} \bat{r_b} & = \Big( \vec{\nu} \cdot \vec{\nabla}_{r, \varphi} \iPsi_j \Big) \mat{r_b} \\ & = \frac{1}{\sqrt{r_b^2 +\dot{r}_b^2}} \Bigg( r_b \pder{\iPsi_j}{r}{} \bat{r_b} -\frac{\dot{r}_b}{r_b} \pder{\iPsi_j}{\varphi}{} \bat{r_b} \Bigg).
		\end{split}
	\label{E_NormalDerivative}
	\end{equation}
Putting this into the second boundary condition~\eqref{E_BoundaryCondition_Two_TE} for TE polarization leads to
	\begin{equation}
		\begin{split}
			0 = \frac{1}{n^2} \pder{\iPsi_1}{r}{} \bat{r_b} -\pder{\iPsi_2}{r}{} \bat{r_b} -\frac{\dot{r}_b}{r_b^2} \Bigg( \frac{1}{n^2} \pder{\iPsi_1}{\varphi}{} \bat{r_b} -\pder{\iPsi_2}{\varphi}{} \bat{r_b} \Bigg).
		\end{split}
	\label{E_BoundaryCondition_Two_TE_Full}
	\end{equation}
At this point there are two options to proceed because the normal derivative~\eqref{E_NormalDerivative} can be expanded as well as the second boundary condition~\eqref{E_BoundaryCondition_Two_TE_Full}. We have checked that both options lead to the same terms in the PT up to second order in $\varepsilon$. We proceed here with condition~\eqref{E_BoundaryCondition_Two_TE_Full} because the first part of this condition equals the boundary condition~\eqref{E_BoundaryCondition_Two_TE_Ge} used by Ge \textit{et al.}~\cite{GSR13} and therefore the corrections can directly be identified. The expansions for the new terms are
	\begin{equation}
		\frac{\dot{r}_b}{r_b^2} = \varepsilon \frac{1}{R} \dot{f} (\varphi) -\varepsilon^2 \frac{2}{R} f (\varphi) \dot{f} (\varphi) +O \big( \varepsilon^3 \big),
	\label{E_CoefficientExpansion}
	\end{equation}
	\begin{equation}
		\begin{split}
			\pder{\iPsi_j}{\varphi}{} \bat{r_b} = & \; \pder{\iPsi_j}{\varphi}{} \bat{R} +\varepsilon R f (\varphi) \mpder{\iPsi_j}{r}{}{\varphi}{}{2} \bat{R} +O \big( \varepsilon^2 \big)
		\end{split}
	\label{E_AngularDerivativeExpansion}
	\end{equation}
and they can be added to the derivation of the PT formulas. The calculation of the corrected PT then follows the same steps as in Refs.~\cite{DBD08,GSR13}, which are unfolded for these additional terms in App.~\ref{A_DerivationOfTheCorrections}.

Importantly, the expansion~\eqref{E_CoefficientExpansion} contains the derivative of the deformation function, $\dot{f} (\varphi)$. We call functions where the maximum of $|\dot{f}(\varphi)|$ is large \textit{strongly winding} boundaries and all others \textit{weakly winding}. For the latter usage of only the radial derivative would produce very small deviations, probably not even recognizable. But for a strongly winding boundary these deviations are relevant, as we will see later. In Ref.~\cite{GSR13} it is not clear if the utilization of only the radial derivative is an intentional simplification or if it was not noticed because only weakly winding boundaries were studied.

The results for the corrected PT are formulated differently from those extracted by Ge \textit{et al.}~\cite{GSR13}. This is done to circumvent the use of derivatives of Bessel and Hankel functions of order higher than the first. To keep the formulas concise the following auxiliary functions have been introduced,
	\begin{equation}
		V_m (x) = \frac{\dot{J}_m^2}{J_m^2} \big( n x \big) -\frac{\dot{H}_m^2}{H_m^2} \big( x \big) +\frac{m^2}{n^2 x^2} \Big( n^2 -1 \Big),
	\label{E_HelpFunction_V}
	\end{equation}
	\begin{equation}
		W_m (x) = n \frac{\dot{J}_m^3}{J_m^3} \big( n x \big) -\frac{\dot{H}_m^3}{H_m^3} \big( x \big) +\frac{3}{2 x} V_m (x) +S_m (x),
	\label{E_HelpFunction_W}
	\end{equation}
	\begin{equation}
		X_m (x) = \frac{\dot{H}_m^2}{H_m^2} \big( x \big) -\frac{m^2}{x^2} +1,
	\label{E_HelpFunction_X}
	\end{equation}
	\begin{equation}
		Y_m (x) = T_m (x) +\frac{m^2}{n^2 x} \Big( n^2 -1 \Big),
	\label{E_HelpFunction_Y}
	\end{equation}
	\begin{equation}
		Z_m (x) = \frac{\dot{H}_m}{H_m} \big( x \big) -\frac{m^2}{x} +x.
	\label{E_HelpFunction_Z}
	\end{equation}
Using these auxiliary functions the PT formulas, in which the corrections are underlined, are
	\begin{widetext}
	\begin{equation}
		b_{1, p} = x_0 S_m A_{mp},
	\label{E_Coefficient_BOne}
	\end{equation}
	\begin{equation}
		\begin{split}
			x_1 = -x_0 \Bigg[ A_{m m} \underbracket{\pm \frac{1}{V_m} \frac{m}{n^2 x_0^2} \Big( n^2 -1 \Big) C_{m m}} \Bigg],
		\end{split}
	\label{E_Eigenvalue_XOne}
	\end{equation}
	\begin{equation}
		a_{1, p} = \frac{x_0}{T_p} \Bigg\{ \Bigg[ \frac{\dot{H}_p}{H_p} S_m +\frac{m^2}{n^2 x_0^2} \Big( n^2 -1 \Big) \Bigg] A_{m p} \underbracket{\pm \frac{m}{n^2 x_0^2} \Big( n^2 -1 \Big) C_{m p}} \Bigg\},
	\label{E_Coefficient_AOne}
	\end{equation}
	\begin{equation}
		\begin{split}
			b_{2, p} = & \, -x_0 \Bigg\{ \Bigg[ \frac{\dot{H}_m^2}{H_m^2} \Big( n^4 -1 \Big) +n^2 -1 \Bigg] x_1 A_{m p} +\frac{1}{2} \bigg[ S_m +x_0 \Big( n^2 -1 \Big) \bigg] B_{m p} \\ & \, -\sum_{q \neq m} S_q a_{1, q} A_{q p} +\sum_{q} \frac{\dot{H}_q}{H_q} b_{1, q} A_{q p} \Bigg\},
		\end{split}
	\label{E_Coefficient_BTwo}
	\end{equation}
	\begin{equation}
		\begin{split}
			x_2 = & \; \frac{1}{V_m} \Bigg[ \Bigg( W_m -\frac{1}{x_0} V_m \Bigg) x_1^2 +X_m x_1 b_{1, m} -\frac{\dot{H}_m}{H_m} b_{2, m} \\ & \, +V_m x_1 A_{m m} -\frac{1}{2} \Bigg( \frac{\dot{H}_m}{H_m} x_0^2 -\frac{3 m^2}{n^2 x_0} \Bigg) \Big( n^2 -1 \Big) B_{m m} -\sum_{q \neq m} Y_q a_{1, q} A_{q m} +\sum_{q} Z_q b_{1, q} A_{q m} \\ & \; \, \underbracket{\pm \frac{2 m}{n^2 x_0} \Big( n^2 -1 \Big) D_{m m} \mp \Big( n^2 -1 \Big) \sum_{q \neq m} \frac{q}{n^2 x_0} a_{1, q} C_{q m} \mp \sum_{q} \frac{q}{x_0} b_{1, q} C_{q m}} \Bigg],
		\end{split}
	\label{E_Eigenvalue_XTwo}
	\end{equation}	
	\begin{equation}
		\begin{split}
			a_{2, p} = & \; \frac{1}{T_p} \Bigg[ V_p x_1 a_{1, p} -X_p x_1 b_{1, p} +\frac{\dot{H}_p}{H_p} b_{2, p} \\ & \, -V_m x_1 A_{m p} +\frac{1}{2} \Bigg( \frac{\dot{H}_m}{H_m} x_0^2 -\frac{3 m^2}{n^2 x_0} \Bigg) \Big( n^2 -1 \Big) B_{m p} +\sum_{q \neq m} Y_q a_{1, q} A_{q p} -\sum_{q} Z_q b_{1, q} A_{q p} \\ & \; \, \underbracket{\mp \frac{2 m}{n^2 x_0} \Big( n^2 -1 \Big) D_{m p} \pm \Big( n^2 -1 \Big) \sum_{q \neq m} \frac{q}{n^2 x_0} a_{1, q} C_{q p} \pm \sum_{q} \frac{q}{x_0} b_{1, q} C_{q p}} \Bigg].
		\end{split}
	\label{E_Coefficient_ATwo}
	\end{equation}
	\end{widetext}
Here, every Bessel, Hankel, and auxiliary function has to be evaluated at $x_0$. The corrections differentiate between the symmetric and antisymmetric wave functions via the signs $\pm$ and $\mp$, with the upper sign for the symmetric wave functions. The corrections can also be identified by the coupling integrals $C_{q p}$ and $D_{q p}$.
Note that $b_{1, p}$ receives no corrections and that $b_{2, p}$ is only influenced by corrections indirectly.

The occurring coupling integrals are the old ones,
	\begin{equation}
		A_{q p} = \frac{c_p}{\pi} \int_{0}^{\pi} f (\varphi) \cos(q \varphi) \cos(p \varphi) d\varphi,
	\label{E_CouplingIntegral_A}
	\end{equation}
	\begin{equation}
		B_{q p} = \frac{c_p}{\pi} \int_{0}^{\pi} f^2 (\varphi) \cos(q \varphi) \cos(p \varphi) d\varphi,
	\label{E_CouplingIntegral_B}
	\end{equation}
	and the two additional integrals
	\begin{equation}
		C_{q p} = \frac{c_p}{\pi} \int_{0}^{\pi} \dot{f} (\varphi) \sin(q \varphi) \cos(p \varphi) d\varphi,
	\label{E_CouplingIntegral_C}
	\end{equation}
	\begin{equation}
		D_{q p} = \frac{c_p}{\pi} \int_{0}^{\pi} f (\varphi) \dot{f} (\varphi) \sin(q \varphi) \cos(p \varphi) d\varphi
	\label{E_CouplingIntegral_D}
	\end{equation}
	that include the deformation function's derivative. The constant is
	\begin{equation}
		c_p = \begin{cases} 1, & p = 0; \\ 2, & p \neq 0, \end{cases}
	\label{E_CouplingIntegralConstant}
	\end{equation}
which arises from the Fourier analysis of the boundary conditions.

By neglecting the latter two coupling integrals~\eqref{E_CouplingIntegral_C} and \eqref{E_CouplingIntegral_D} the formulas yield the uncorrected PT, but if the maximum absolute value of the deformation function's derivative is large these integrals cannot be neglected. Note that the indices of the coupling integrals are interchanged, in comparison to Ref.~\cite{GSR13}, to match the derivation procedure of the PT. This way the last index belongs to the Fourier order with which the expanded boundary conditions have been analyzed to reach the corresponding frequencies and coefficients. Also note the mixed occurrences of sines and cosines, which result from the derivatives with respect to $\varphi$. The integrand, however, always remains symmetric because for a symmetric deformation function $f (\varphi)$ the derivative $\dot{f} (\varphi)$ is antisymmetric and negates the antisymmetry of every sine.

For the antisymmetric wave functions, as in the previous perturbation theories~\cite{DBD08,GSR13}, all sines and cosines have to be switched while setting $c_p = 0$ for $p = 0$.

\section{Applicability of the perturbation series}
\label{sec:applicabilit}
In this section we discuss the validity of the perturbation series. Closely following the discussion for the TM PT in Ref.~\cite{DBD08}, our criterium for the validity is that the change of the wave function inside the cavity is small in first-order of the perturbation, 
\begin{equation}\label{eq:c}
\varepsilon|a_{1,p}| \ll 1 
\end{equation}
with $p\neq m$ and $a_{1,p}$ from Eq.~(\ref{E_Coefficient_AOne}). The deformation strength $\varepsilon$ is here chosen to be nonnegative without loss of generality. The condition~(\ref{eq:c}) holds if the two following conditions are simultaneously fulfilled
\begin{equation}\label{eq:cA}
\varepsilon \Bigg[ \meanb{\left|\frac{1}{T_p}\frac{\dot{H}_p}{H_p} S_m\right|} x_0 + \meanb{\left|\frac{1}{T_p}\right|}\frac{m^2}{n^2 x_0} \Big( n^2 -1 \Big) \Big] \left|A_{m p}\right| \ll 1 \ , 
\end{equation}
\begin{equation}\label{eq:cC}
\varepsilon \meanb{\left|\frac{1}{T_p}\right|}\frac{m}{n^2 x_0} \Big( n^2 -1 \Big) \left|C_{m p}\right| \ll 1 \ ,
\end{equation}
where $\mean{F_p}$ indicates the typical value of $F_p$. The small imaginary part of $x_0$ is ignored here. 
The conditions (\ref{eq:cA}) and~(\ref{eq:cC}) are challenged by angular momenta $p$ with $T_p(x_0) \approx 0$. In this case we approximate
\begin{equation}
\meanb{\left|\frac{1}{T_p}\right|} = \left|\frac{1}{\dot{T}_p\delta x}\right|
\end{equation}
with 
\begin{equation}
\dot{T}_p(x_0) = -\frac{(n^2-1)p^2}{n^2x_0^2}-\frac{\dot{H}_p}{H_p} S_p
\end{equation}
as derived in Ref.~\cite{GSR13} for $T_p = 0$. In that case $S_p$ can be replaced by $(n^2-1)\dot{H}_p/H_p$ by exploiting Eqs.~(\ref{eq:Sm})-(\ref{eq:Tm}). The typical distance between two frequencies is estimated using Weyl's law in Ref.~\cite{DBD08} 
\begin{equation}
\delta x = \frac{4}{n^2s_nx_0}
\end{equation}
with
\begin{equation}
s_n = 1-\frac{2}{\pi}\left(\arcsin{\frac{1}{n}}+\frac{1}{n}\sqrt{1-\frac{1}{n^2}}\right) \ .
\end{equation}

Using the asymptotic behaviour of WGMs with azimuthal mode number $q$~\cite{DBD08}
\begin{equation}\label{eq:x0mn}
x_0 = \frac{q}{n} + {\cal O}\left(q^{1/3}\right) 
\end{equation}
and the asymptotic of the Hankel function~\cite{BD12}
\begin{equation}
\frac{\dot{H}_q}{H_q}(x_0) = -\sqrt{\frac{q^2}{x_0^2}-1} + {\cal O}\left(\frac{1}{x_0}\right)
\end{equation}
we find the two criteria for the applicability of the perturbation series
\begin{equation}\label{eq:cA2}
\varepsilon x_0^2n^2 \frac{s_n}{4}|A_{m p}| \ll 1 \ ,
\end{equation}
\begin{equation}\label{eq:cC2}
\varepsilon \frac{m^2}{n^2} \frac{s_n}{4}\frac{|C_{m p}|}{m} \ll 1 \ .
\end{equation}
The criterium~(\ref{eq:cA2}) equals the one for the TM PT, which, however, was further processed in Ref.~\cite{DBD08} by introducing the perturbation area, that is the area of the region where the refractive index differs from the one of the circular cavity, see Fig.~\ref{F_MicroflowerLimacon_Boundaries}. 

At first glance, it might appear strange that condition~(\ref{eq:cA2}) depends on $x_0$ but condition~(\ref{eq:cC2}) does not. However, Eq.~(\ref{eq:x0mn}) shows that also condition~(\ref{eq:cC2}) depends on $x_0$.

\section{Numerical results}
\label{S_ResultsForTheCorrectedTheory}
\begin{figure}
		\includegraphics[width=0.96\columnwidth]{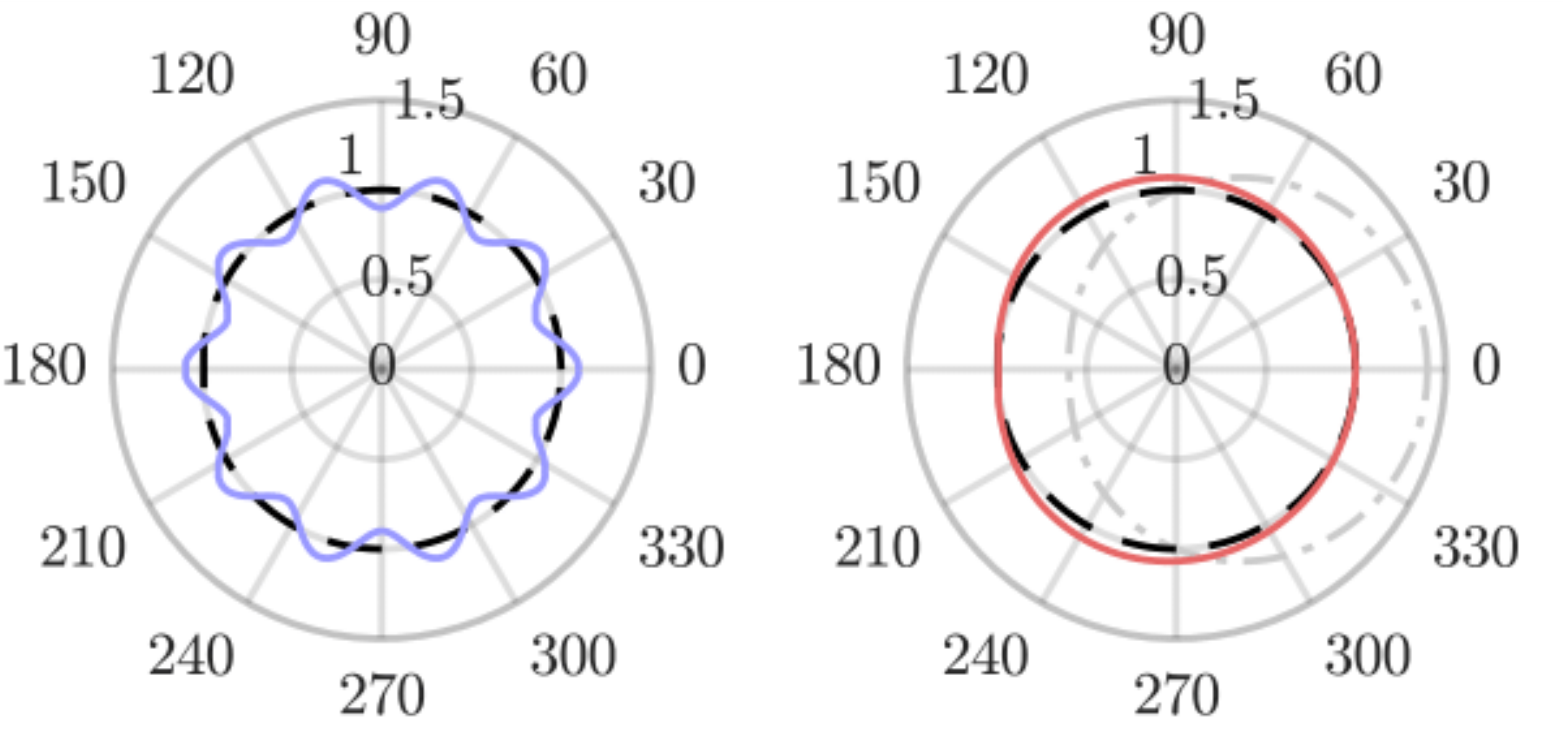}
		\caption{Boundary of the microflower cavity (left) with $\kappa = 10$ and $\varepsilon = 0.1$ and the lima\c{c}on cavity (right), including the regular (dash-dotted light gray) and shifted (red/dark gray) boundary, with $\varepsilon = 0.4$. The unperturbed circular cavity is shown as the dashed curve. The radial coordinate is here dimensionless as $R = 1$ is chosen.}
	\label{F_MicroflowerLimacon_Boundaries}
	\end{figure}

The microflower cavity is a clear case to show the improvements brought by the corrections, but the lima\c{c}on cavity profits as well. Both are described by the deformation function
	\begin{equation}
		f (\varphi) = \cos(\kappa \varphi)
	\label{E_Microflower_DeformationFunction}
	\end{equation}
with the deformation parameter $\kappa$. The boundary of the cavity is then given by $r (\varphi) = R \big[ 1 +\varepsilon \cos(\kappa \varphi) \big]$. While the lima\c{c}on cavity is defined by $\kappa = 1$, let $\kappa \ge 3$ define the microflower cavity. In Fig.~\ref{F_MicroflowerLimacon_Boundaries} both cavities are illustrated.
The derivative of the deformation function is
	\begin{equation}
		\dot{f} (\varphi) = -\kappa \sin(\kappa \varphi)
	\end{equation}
which is, roughly speaking, $\kappa$ times larger than the deformation function itself. This implies that the coupling integrals $C_{q p}$ and $D_{q p}$ mostly are $\kappa$ times larger than $A_{q p}$ and $D_{q p}$, too. Hence, for large $\kappa$ the microflower cavity qualifies as a cavity with a strongly winding boundary.

With the maximum values of $|A_{m p}| = 1/2$ and $|C_{m p}| = \kappa/2$ the criteria for the applicability of the perturbation series~(\ref{eq:cA2})-(\ref{eq:cC2}) can be written as
\begin{equation}\label{eq:cA2flower}
\varepsilon \ll \frac{8}{x_0^2n^2 s_n}  \ ,
\end{equation}
\begin{equation}\label{eq:cC2flower}
\varepsilon \ll \frac{8n^2}{m\kappa s_n} \ .
\end{equation}

\subsection{Microflower cavity}
\label{SS_Microflower}
Boriskina \textit{et al.} introduced the microflower cavity in Ref.~\cite{BBS03} as a smooth version of the microgear cavity, which is a disk having a grating of period $\kappa$. This particular boundary modification enhances the $Q$-factor of the even-parity TE-polarized mode with the azimuthal mode number $m = \kappa/2$ and spoils the $Q$-factor of all other modes~\cite{FB01,FB02}. The microgear and the microflower are therefore ideal geometries for mode selection in lasers.  

The deformation parameter used in our studies is $\kappa = 10$ and therefore the relevant azimuthal mode number is $m = 5$. Figure~\ref{F_Microflower_TE_M5_13_Intensities} shows an example computed from the boundary element method (BEM)~\cite{Wiersig02b} using $10\,000$ discretization points along the cavity's boundary. It can be seen that the even-parity mode fits into the cavity very well while the odd-parity mode is strongly mismatched. The result is a significant splitting of the real and imaginary parts, and in turn the $Q$-factor
	\begin{equation}
		Q = -\frac{\text{Re}(x)}{2 \text{Im}(x)}
	\label{E_QFactor}
	\end{equation}
of the modes. While the origin of this splitting is intuitively clear from the mode pattern, the unexpected observation is that the $Q$-factor of the even-parity mode increases for small deformations. As stronger deformations reduce a $Q$-factor in the end, a maximum of the $Q$-factor occurs for the even-parity mode. {For the microgear cavity, this $Q$-factor enhancement was predicted in Ref.~\cite{FB01} and experimentally confirmed in Ref.~\cite{FB02}. It has been explained in terms of the discontinuous boundary condition~\cite{FB01,HMB05}.
%
	\begin{figure}
		\includegraphics[width=0.96\columnwidth]{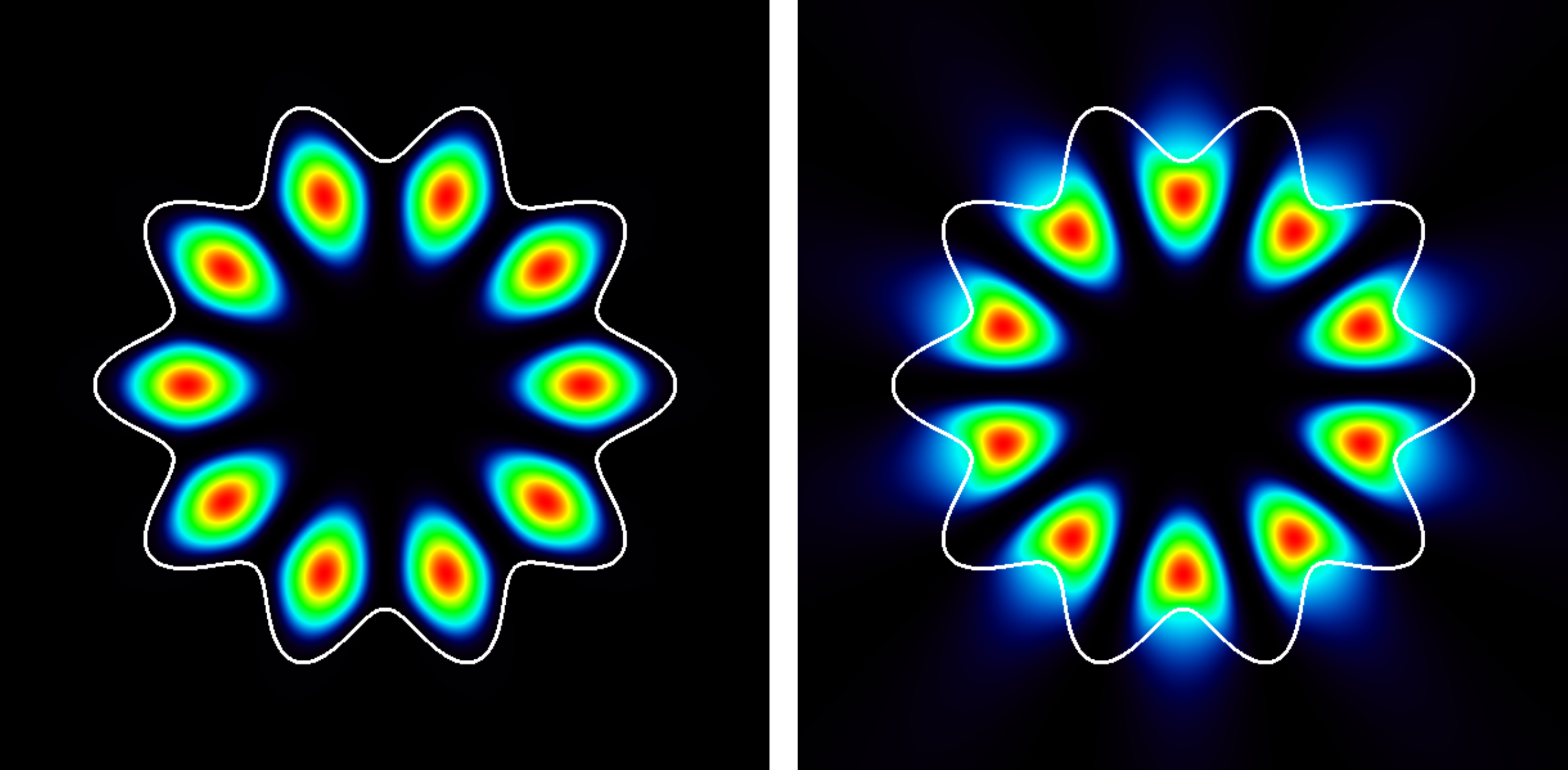}
		\caption{Microflower cavity with $n = 2.63$, $\kappa = 10$, $\varepsilon = 0.13$: intensity distributions of the even-parity (left) and odd-parity mode (right) with the mode numbers $l = 1$ and $m = 5$ calculated by the BEM; the deformation strength $\varepsilon$ is rather large because it is taken at the maximum $Q$-factor according to the BEM results in Fig.~\ref{F_Microflower_TE_M5_QFactor}.}
	\label{F_Microflower_TE_M5_13_Intensities}
	\end{figure}

Figures~\ref{F_Microflower_TE_M5_RealImaginary} and~\ref{F_Microflower_TE_M5_QFactor} reveal that the uncorrected PT cannot reproduce the above phenomenon. In particular, the splitting in the imaginary part of the frequency and the $Q$-factor is too small if compared to the full numerical simulations based on the BEM. Furthermore, in the uncorrected PT both parities split up falsely since the \textit{even} parity has the \textit{lower} $Q$-factor. This means that the principally correct direction of the real and imaginary part's splitting does not translate to the $Q$-factor. The reason for this behavior has not been analyzed further but it has been found in the TM PT as well.
	\begin{figure}
		\includegraphics[width=0.96\columnwidth]{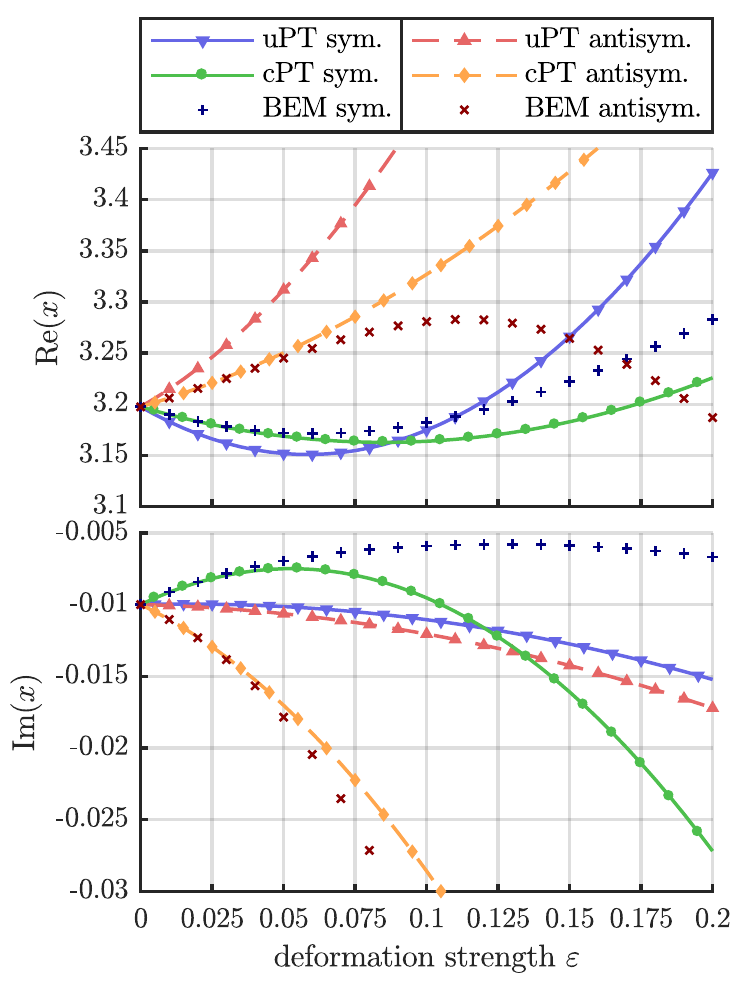}
		\caption{Microflower cavity with $n = 2.63$, $\kappa = 10$: real (top) and imaginary parts (bottom) of the dimensionless frequency in dependence on the dimensionless deformation strength. Solid (dashed) curves show the results for the even-parity (odd-parity) mode with the mode numbers $l = 1$ and $m = 5$. The triangles (circle/diamond) correspond to the uncorrected (corrected) perturbation theory. Symbols without curves correspond to the BEM.}
	\label{F_Microflower_TE_M5_RealImaginary}
	\end{figure}
	\begin{figure}
		\includegraphics[width=0.96\columnwidth]{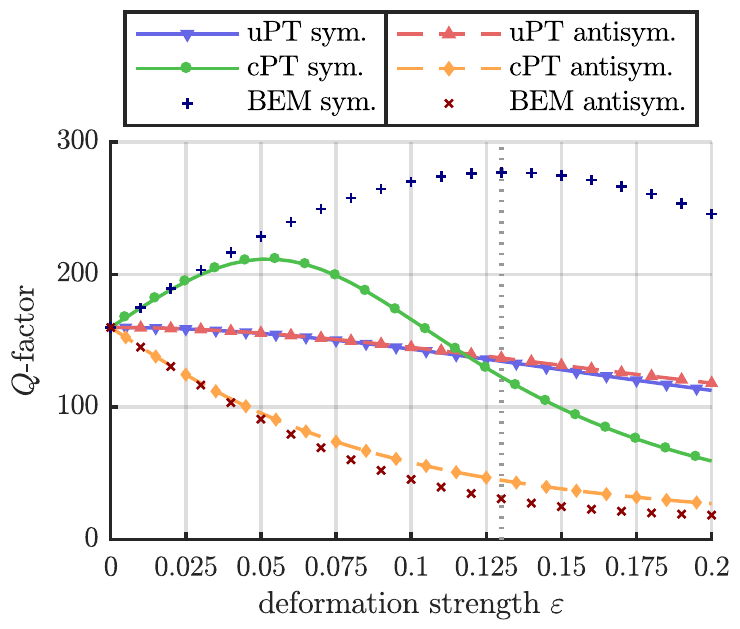}
		\caption{Microflower cavity with $n = 2.63$, $\kappa = 10$, $l = 1$, $m = 5$: $Q$-factor versus dimensionless deformation strength; the dotted vertical line marks the maximum $Q$-factor according to the BEM.}
	\label{F_Microflower_TE_M5_QFactor}
	\end{figure}

The corrections introduced in Sec.~\ref{S_CorrectionOfTheTETheory} solve these problems as can be seen in Figs.~\ref{F_Microflower_TE_M5_RealImaginary} and \ref{F_Microflower_TE_M5_QFactor} as well. The splitting between both parities is very accurately described up to a deformation strength of $\varepsilon \approx 0.03$, which is henceforth considered the PT's regime of applicability for the given cavity parameters. A maximum in the even parity's curve lies at around $\varepsilon \approx 0.05$ while the antisymmetric one is only descending. This maximum is not at the same position as the one of the BEM but it is clearly visible. 

An agreement with the BEM up to its maximum cannot be expected here because such deformations are probably excluded by the first condition~(\ref{eq:cA2flower}) with $\varepsilon \ll 0.21$. Surprisingly, the second condition~(\ref{eq:cC2flower}) with $\varepsilon \ll 2.1$ is much weaker for the present cavity. The validity of the Rayleigh hypothesis~\cite{BergFokkema79} with $\varepsilon \approx 0.046$, cf. Sec.~\ref{S_PerturbationTheories}, is here a strong bound. It gives a good prediction of the critical deformation at which the PT starts to fail.
	\begin{figure}
		\includegraphics[width=0.96\columnwidth]{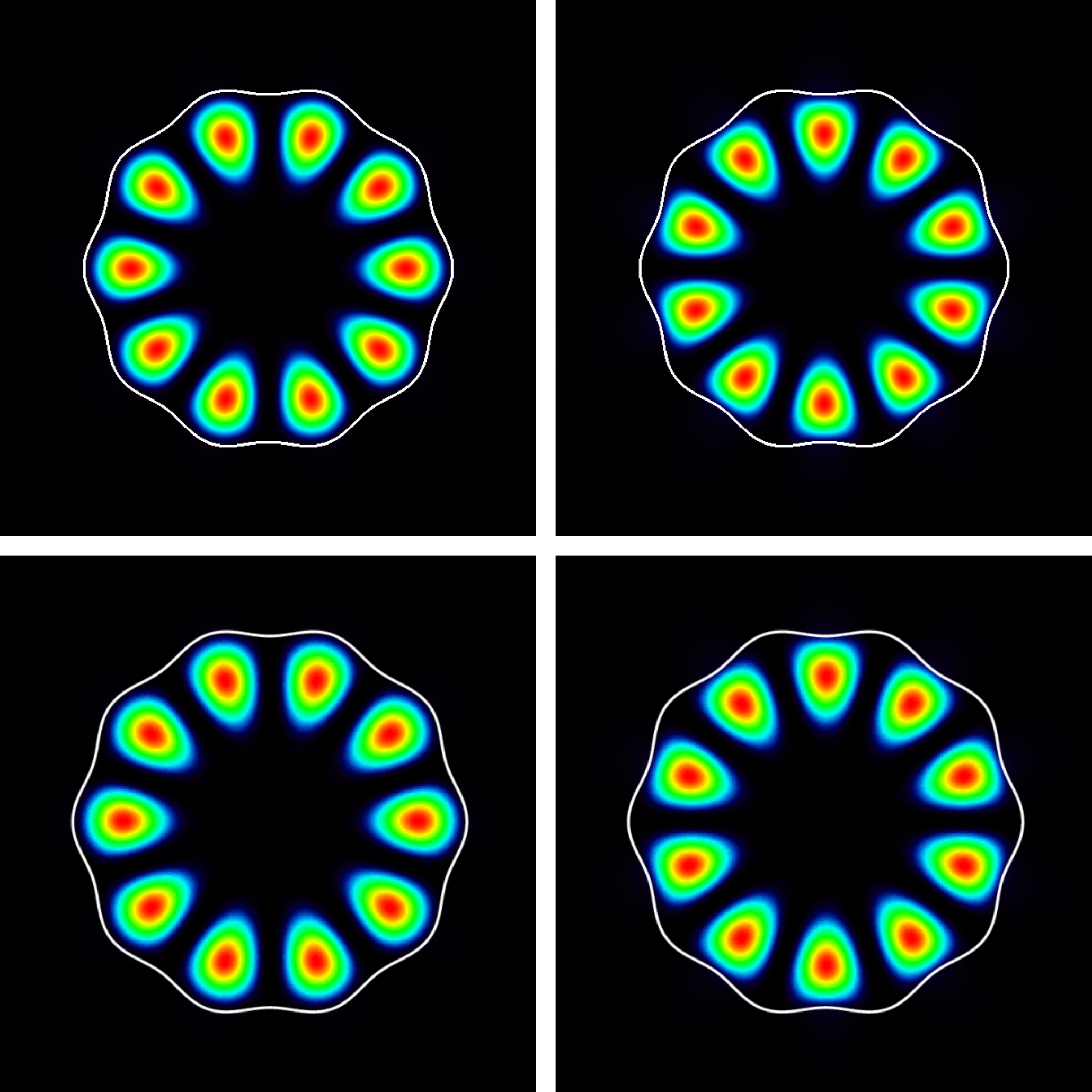}
		\caption{Microflower cavity with $n = 2.63$, $\kappa = 10$, $\varepsilon = 0.03$: intensity distributions of the even-parity (left) and odd-parity mode (right) with the mode numbers $l = 1$ and $m = 5$ for the BEM (top) and corrected PT (bottom).}
	\label{F_Microflower_TE_BEM_M5_03_Intensities}
	\end{figure}

The difference between the PTs manifests in the formulas of their frequencies. Using $A_{m m} = \pm \delta_{2 m, \kappa}/2$, the \textit{uncorrected} PT's frequencies $\tilde{x}$ can be written as
	\begin{equation}
		\tilde{x} = x_0 \mp \varepsilon \frac{x_0}{2} \delta_{2 m, \kappa} +\varepsilon^2 C \big( x_0^2 \big) +O \big( \varepsilon^3 \big),
	\label{E_EigenvalueShort_Uncorrected}
	\end{equation}
where $C \big( x_0^2 \big)$ is a constant independent of the parity. Using $C_{m m} = -\kappa \delta_{2 m, \kappa}/2$ an analogous formula is attainable for the \textit{corrected} PT, but with an essential difference in the first order,
	\begin{equation}
		\begin{split}
			x = & \; \, x_0 \mp \varepsilon \frac{x_0}{2} \delta_{2 m, \kappa} \Bigg[ 1 -\frac{1}{V_m} \frac{m \kappa}{n^2 x_0^2} \Big( n^2 -1 \Big) \Bigg] \\ & \, +\varepsilon^2 C \big( x_0^2 \big) +O \big( \varepsilon^3 \big).
		\end{split}
	\label{E_EigenvalueShort_Corrected}
	\end{equation}
The second order gets corrected too, but there is no difference between both parities, which leaves the first order as the most important contributor to the splitting.

As can be seen in Eq.~\eqref{E_EigenvalueShort_Uncorrected} the factor $1/2$ of the uncorrected theory's first order does \textit{not} distinguish the real and imaginary parts of $x_0$ because it is real. In the corrected theory this is different. The factor in Eq.~\eqref{E_EigenvalueShort_Corrected} is a {\em complex number} because of $V_m (x_0)$ and $x_0^2$ in the denominator. Remember that this complex modification exists solely because of the angular derivative in Eq.~(\ref{E_BoundaryCondition_Two_TE_Full}) that couples to the boundary function's derivative. That complex factor implies that real and imaginary parts can be modified quite differently. For example with $n = 2.63$, $\kappa = 10$, and $m = 5$ the first order of the frequency for the uncorrected and the corrected PT, respectively, is
	\begin{equation*}
		\begin{split}
			& \tilde{x}_1 = \mp \Big( 1.5988 -0.0050i \Big), \\ & x_1 = \mp \Big( 0.8152 -0.0953i \Big).
		\end{split}
	\end{equation*}
	The real part of the corrected frequency is about $1/2$ of the uncorrected while the imaginary part is $20$ times bigger. This considerable difference results in a much larger splitting of the imaginary parts and likewise the $Q$-factors for small $\varepsilon$.

The intensity distributions that can be calculated by the PTs do \textit{not} show a significant difference. In the relevant regime of small deformation strength they are nearly indistinguishable. For the deformation strength of $\varepsilon = 0.03$ the distributions are plotted in Fig.~\ref{F_Microflower_TE_BEM_M5_03_Intensities} for the BEM and the corrected PT. Both agree well, as does the uncorrected PT (not shown). Only for considerably larger deformation strength differences become visible, for the antisymmetric wave functions first. In small regions close to the boundary the intensity starts to deviate from their surroundings and these regions grow with increasing deformation strength. The intensity overshoots the regular amplitude between deformation strengths of $\varepsilon = 0.06$ and $\varepsilon = 0.14$ and rapidly diverges thereafter.

To present results for modes with different azimuthal mode numbers $m$ it is convenient to consider the real and imaginary parts of the differences between the PTs and the BEM, $\Delta x = x_{\text{PT}} -x_{\text{BEM}}$. Figure~\ref{F_Microflower_TE_03_RealImaginaryDiffernece} shows the differences for the mode numbers $m = 5, \ldots, 12$ and fixed perturbation strength $\varepsilon = 0.03$. For each $m$ the deformation parameter $\kappa$ is adjusted to be equal to $2m$ because the long-lived mode resulting from the mode selection is the most interesting one.  For those modes the real parts rise in the interval Re$(x) \approx 3, ..., 6.5$ and the imaginary parts decay in Im$(x) \approx -0.014, ..., -0.03 \cdot 10^{-4}$. Nevertheless, the relative error with respect to the BEM rises with increasing $m$. Therefore, the lower panel of Fig.~\ref{F_Microflower_TE_03_RealImaginaryDiffernece} does \textit{not} imply that the imaginary parts of the PTs get more accurate. Instead, it shows that in this case the corrected PT always outperforms the uncorrected PT.
\begin{figure}
	\includegraphics[width=0.96\columnwidth]{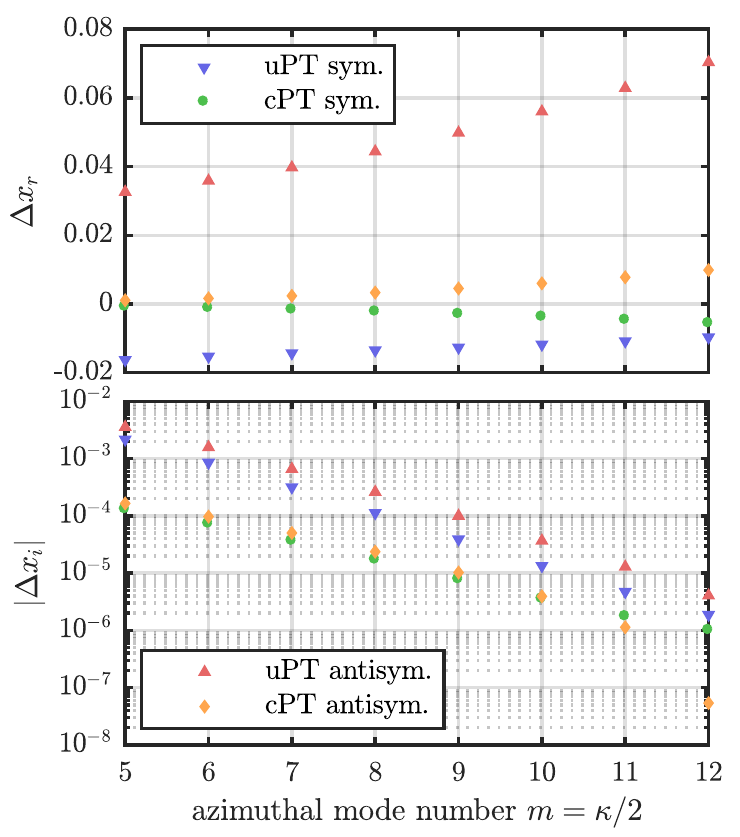}
	\caption{Microflower cavity with $n = 2.63$, $\varepsilon = 0.03$, and $l = 1$: dimensionless deviation of the two PT's real (top) and imaginary (bottom, on a logarithmic scale) parts of the frequency from the BEM reference $\Delta x = x_{\text{PT}} -x_{\text{BEM}}$ versus the azimuthal mode number $m = \kappa/2$.}
	\label{F_Microflower_TE_03_RealImaginaryDiffernece}
\end{figure}

\subsection{Lima\c{c}on cavity}
\label{SS_Limacon}
The lima\c{c}on cavity was studied using the shifting method which was introduced in Ref.~\cite{KW14} for the TM PT. The horizontal shifting by $-\varepsilon$ places the lima\c{c}on cavity centered at the origin which minimizes the perturbation area, see Fig.~\ref{F_MicroflowerLimacon_Boundaries}(b). A reduced perturbation area, in turn, increases the accuracy of the PT. With the shifting method the lima\c{c}on cavity can very well be identified as a disk with a weakly winding boundary, but even here improvements induced by the corrected PT manifest. 

For the lima\c{c}on cavity we consider the real and imaginary parts of the difference $\Delta x$ as well. Figure~\ref{F_Limacon_TE_M4_RealDifferences} shows the real part of the frequency in a linear and a semi-logarithmic scale. It can be observed that the corrected PT starts off with a bigger deviation if compared to the uncorrected PT but gets more accurate after around $\varepsilon = 0.15$. It has to be noted that the existence and position of the downward dips, that tell where the PT results cross the BEM references, depends on various parameters. For the uncorrected PT the mode numbers, the deformation and its strength decide, via the coupling integrals, if and where a crossing takes place. The shift is also relevant, it can completely change the behavior of the PT. In this regard, $-\varepsilon$ is not generally the best value for the shift, which depends on the parameters mentioned before as well. In our experience it is mostly chance, or at least not systematic, if the PT crosses the BEM references or not.
	\begin{figure}
		\includegraphics[width=0.96\columnwidth]{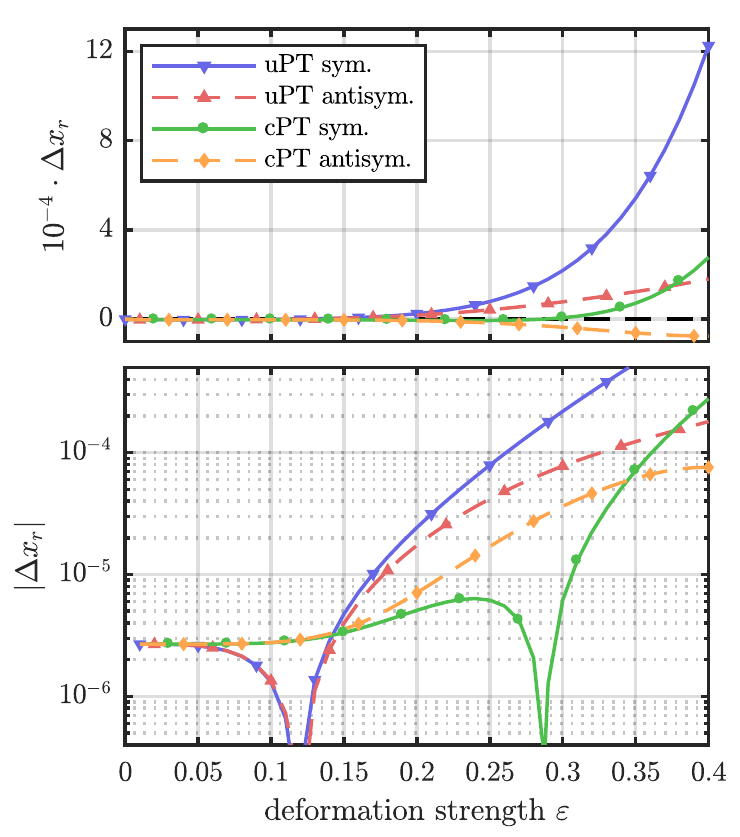}
		\caption{Lima\c{c}on cavity with $n = 2.63$, $l = 1$, and $m = 4$: dimensionless deviation of the two PT's real parts of the frequency from the BEM reference in dependence on the dimensionless deformation strength; the lower panel shows the same data on a logarithmic $y$-axis; the dips appear where the PT results cross those of the BEM.}
	\label{F_Limacon_TE_M4_RealDifferences}
	\end{figure}

Figure~\ref{F_Limacon_TE_M4_ImaginaryDifferences} shows the imaginary part of the frequency again in a linear and a semi-logarithmic scale. It can be observed that the corrected PT is more accurate right from the beginning with a considerable lead between $\varepsilon = 0.1$ and $\varepsilon = 0.2$. For even higher deformation strengths the corrected and the uncorrected PTs start to align. As for the microflower cavity, the imaginary part is the more relevant benchmark because it is generally $1$ or $2$ orders of magnitude smaller than the real part and deviations of the same magnitude weigh heavier on it.
	\begin{figure}
		\includegraphics[width=0.96\columnwidth]{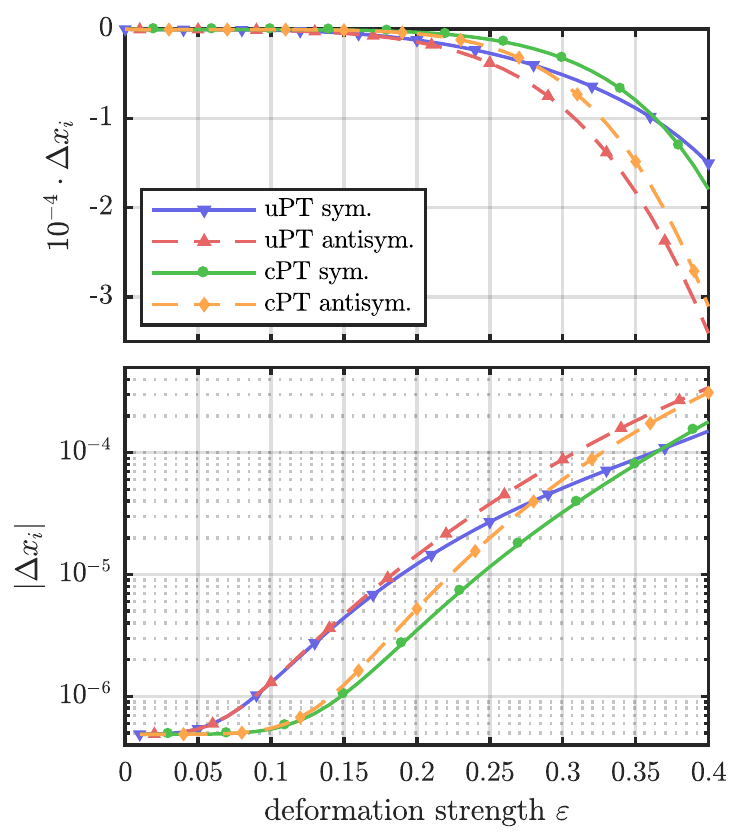}
		\caption{Lima\c{c}on cavity with $n = 2.63$, $l = 1$, and $m = 4$: dimensionless deviation of the two PT's imaginary parts of the frequency from the BEM reference in dependence on the dimensionless deformation strength; the lower panel shows the same data on a logarithmic $y$-axis.}
	\label{F_Limacon_TE_M4_ImaginaryDifferences}
	\end{figure}

Overall, it can be said that the corrections improve the PT's results. Note that for small $\varepsilon$ the differences between the BEM and the PTs do not tend to zero because for very small $\varepsilon$ the BEM suffers from the finite discretization mesh of the cavity's boundary.

The first condition~(\ref{eq:cA2}) evaluated with the perturbation area introduced in Ref.~\cite{DBD08} gives $\varepsilon \approx 0.25$. However, we observe that the PT works very well up to $\varepsilon \approx 0.32$. This is the value up to which the difference to the BEM result is smaller than $10^{-4}$, see Fig.~\ref{F_Limacon_TE_M4_ImaginaryDifferences}. The second condition~(\ref{eq:cC2}) is expected to be even more irrelevant than in the microflower cavity as $\kappa = 1$. The criterion of the Rayleigh hypothesis~\cite{BergFokkema79} with $\varepsilon \approx 2/3$, see Sec.~\ref{S_PerturbationTheories}, overestimates the range of applicability.

A comparison with the TM theory is shown in Fig.~\ref{F_Limacon_TETM_M4_RealImaginaryDifference}. While the real parts of the corrected TE theory start of slightly worse, they get more accurate than the TM theory's around the deformation strength $\varepsilon = 0.15$. In the imaginary parts the theories are well comparable with a slight lead of the TE PT.
	\begin{figure}
		\includegraphics[width=0.96\columnwidth]{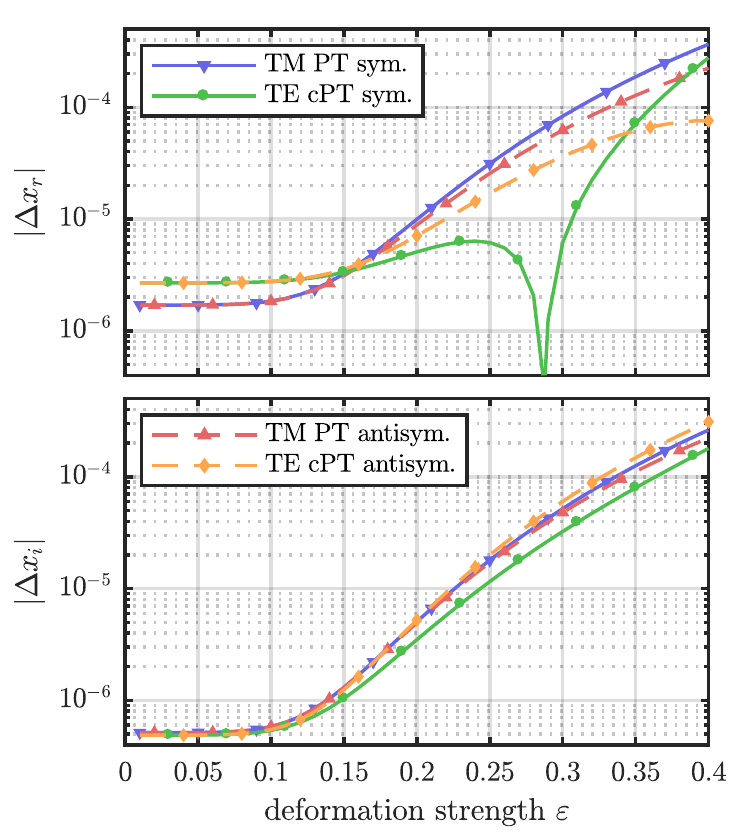}
		\caption{Lima\c{c}on cavity with $n = 2.63$, $l = 1$, and $m = 4$: dimensionless deviation of the TM PT and the corrected TE PT real (top) and imaginary parts (bottom) of the frequency from the BEM reference versus the dimensionless deformation strength in semi-logarithmic plots; the dip appears where the PT results cross those of the BEM.}
	\label{F_Limacon_TETM_M4_RealImaginaryDifference}
	\end{figure}

\section{Conclusion}
\label{S_Conclusion}
The perturbation theory for TE whispering-gallery modes by Ge \textit{et al.}~\cite{GSR13} has been corrected by including the full spatial variation of the normal derivative along the boundary of the cavity. The obtained formulas accurately describe the microflower cavity with its strongly winding boundary up to a deformation strength of $\varepsilon \approx 0.03$. This is consistent with the derived regime of applicability of the perturbation series and is close to the rough estimation of $\varepsilon \approx 0.046$ for a microflower depending on the Rayleigh hypothesis~\cite{BergFokkema79}. 
While the uncorrected perturbation theory exhibits a much too small splitting of the imaginary part of the frequency and the $Q$-factor, the splitting of the corrected theory coincides very well with full numerical results. Most importantly, the corrected theory can describe the $Q$-factor enhancement, which allows for a mode selection without the damaging effect of $Q$-spoiling.

Even for the lima\c{c}on cavity with its weakly winding boundary the results improve, although the normal and radial derivatives do not deviate much up to quite large deformation strengths. While the real part profits for higher deformation strengths only, more importantly, the imaginary parts receive a higher accuracy of one order of magnitude over most of the studied regime.

Overall, the corrected TE perturbation theory can be used to examine cavities with strongly winding boundaries and slight improvements can be expected for weakly winding ones. In addition, the corrected TE theory, in contrast to the uncorrected one, provides results as good as the TM theory. The corrected theory requires the calculation of two additional coupling matrices, which include the derivative of the deformation function, leading to an approximately doubled calculation time. Compared to the BEM or similar numerical methods, however, it remains a much faster alternative for small deformations.

\begin{acknowledgments}
\label{Acknowledgments}
The authors would like to thank J. Kullig, S. Neumeier, and L. Ge for discussions. Financial support by the Graduiertenf\"{o}rderung des Landes Sachsen-Anhalt and the DFG (project WI1986/7-1) is acknowledged.
\end{acknowledgments}

\appendix
\section{Derivation of the Corrections}
\label{A_DerivationOfTheCorrections}
The procedure to derive the PT formulas, as outlined first by Dubertrand \textit{et al.} in Ref.~\cite{DBD08}, is applied in this appendix to extract the correction terms for the TE theory. 
The overall goal is to expand the boundary conditions~\eqref{E_BoundaryCondition_One} and \eqref{E_BoundaryCondition_Two_TE_Full} at $\varepsilon = 0$ in several intermediate steps and to perform a Fourier analysis. The first step is to expand the derivatives in both boundary conditions, and in our case the additional quotient $\dot{r}_b/r_b^2$, because $r_b$ depends on $\varepsilon$. Here, the expansions needed are in Eqs.~\eqref{E_CoefficientExpansion} and \eqref{E_AngularDerivativeExpansion} which have to be inserted into the additional terms in the second boundary condition~\eqref{E_BoundaryCondition_Two_TE_Full}, leading to
	\begin{widetext}
	\begin{equation}
		\begin{split}
			& -\frac{\dot{r}_b}{r_b^2} \Bigg( \frac{1}{n^2} \pder{\iPsi_1}{\varphi}{} \bat{r_b} -\pder{\iPsi_2}{\varphi}{} \bat{r_b} \Bigg) = \\ & \; \; \; \; \; \; \; \, -\Bigg[ \varepsilon \frac{1}{R} \dot{f} (\varphi) -\varepsilon^2 \frac{2}{R} f (\varphi) \dot{f} (\varphi) \Bigg] \Bigg[ \frac{1}{n^2} \pder{\iPsi_1}{\varphi}{} \bat{R} -\pder{\iPsi_2}{\varphi}{} \bat{R} +\varepsilon R f (\varphi) \Bigg( \frac{1}{n^2} \mpder{\iPsi_1}{r}{}{\varphi}{}{2} \bat{R} -\mpder{\iPsi_2}{r}{}{\varphi}{}{2} \bat{R} \Bigg) \Bigg] +O \big( \varepsilon^3 \big).
		\end{split}
	\label{E_BoundaryCondition_Two_TE_Correction_Expanded}
	\end{equation}
	\end{widetext}
Because we expand up to the second order in $\varepsilon$, every higher order term can be ignored.

To continue, the ansatz~\eqref{E_DeformationWaveFunctions} is inserted into Eq.~\eqref{E_BoundaryCondition_Two_TE_Correction_Expanded} and the derivatives are calculated; remember to apply the chain rule. The derivative with respect to $\varphi$ simply produces terms containing $-p \sin(p \varphi)$, this means that in the first two derivatives the Bessel and Hankel functions cancel at $r = R$. The derivative with respect to $r$ only works on the Bessel and Hankel functions which produces the factor $nk$ for the Bessel and $k$ for the Hankel functions. The explicit presentation of the Bessel and Hankel functions derivative does not need to be used because the PT uses $\dot{J}_p/J_p$ and $\dot{H}_p/H_p$ as standard elements to keep the formulas as compact as possible. The result is
	\begin{widetext}
	\begin{equation}
		\begin{split}
			& -\frac{\dot{r}_b}{r_b^2} \Bigg( \frac{1}{n^2} \pder{\iPsi_1}{\varphi}{} \bat{r_b} -\pder{\iPsi_2}{\varphi}{} \bat{r_b} \Bigg) =  \\ & \; \; \; \; \; \; \; \, -\Bigg\{ \underbrace{\varepsilon \frac{1}{R} \dot{f} (\varphi)}_{I} \underbrace{-\varepsilon^2 \frac{2}{R} f (\varphi) \dot{f} (\varphi)}_{II} \Bigg\} \Bigg\{ \underbrace{-\bigg( \frac{1}{n^2} -1 \bigg) m \sin(m \varphi)}_{III} \underbrace{-\sum_{p \neq m} \bigg( \frac{1}{n^2} -1 \bigg) p a_p \sin(p \varphi) +\sum_{p} p b_p \sin(p \varphi)}_{IV} \\ & \; \; \; \; \; \; \; \, \underbrace{-\varepsilon k R f (\varphi)}_{V} \Bigg[ \underbrace{\Bigg( \frac{1}{n} \frac{\dot{J}_m}{J_m} \bat{x} -\frac{\dot{H}_m}{H_m} \bat{x} \Bigg) m \sin(m \varphi)}_{VI} \underbrace{+\sum_{p \neq m} \Bigg( \frac{1}{n} \frac{\dot{J}_p}{J_p} \bat{x} -\frac{\dot{H}_p}{H_p} \bat{x} \Bigg) p a_p \sin(p \varphi) -\sum_{p} \frac{\dot{H}_p}{H_p} \bat{x} p b_p \sin(p \varphi)}_{VII} \Bigg] \Bigg\}
		\end{split}
	\end{equation}
	\end{widetext}
	where the argument is $x (\varepsilon) = k (\varepsilon) R$, see Eq.~\eqref{E_EigenvalueExpansion}, which has to be considered for the next expansion.

At this state of an expansion terms of third and higher orders can be sorted out for the first time. In this case the argumentation is as follows. (a) Term $III$ is a zeroth-order term and generates a first-order term in combination with term $I$ as well as a second-order term in combination with term $II$. (b) Term $IV$ has first- and second-order terms because of the coefficients $a_p$ and $b_p$, see Eqs.~\eqref{E_CoefficientAExpansion} and \eqref{E_CoefficientBExpansion}. Therefore, it can be combined with term $I$ to a second-order term by keeping $a_{1, p}$ and $b_{1, p}$ but discarding $a_{2, p}$ and $b_{2, p}$. The combination of terms $IV$ and $II$ can be discarded as well, because it is of third and higher order. (c) Terms $I$ and $V$ generate a second-order term so the factor $k R$ in term $V$ and the term $VI$ have to be accounted for in zeroth order, meaning $k R \rightarrow x_0$ and the argument of the Bessel and Hankel functions has to be $x \rightarrow x_0$ too, see Eq.~\eqref{E_EigenvalueExpansion}. Term $VI$ then equals the conditional equation~\eqref{E_ConditionalEquation_TE} with $T_m \at{x_0} = 0$, and it therefore vanishes. (d) Term $VII$ is, just as term $IV$, of first and second order and vanishes in the third/fourth order when combined with terms $I$ and $V$ or the fourth/fifth order in combination with terms $II$ and $V$. All in all, because of (c) and (d) the terms $V$, $VI$, and $VII$ are irrelevant.

Note that this simplifies the corrections a lot because no $x (\varepsilon)$ and no Bessel and Hankel functions remain in the formulas. If those would still be present, the expansion~\eqref{E_EigenvalueExpansion} would have to be inserted for $x (\varepsilon)$ instead, and more laborious, the Bessel and Hankel functions would have to be expanded at $\varepsilon = 0$. The latter would have included the denominator, led to higher derivations of the Bessel and Hankel functions and resulted in much more terms, as it does in the formulas of the uncorrected PT.
This way the remaining expansion is just
	\begin{widetext}
	\begin{equation}
		\begin{split}
			& -\frac{\dot{r}_b}{r_b^2} \Bigg( \frac{1}{n^2} \pder{\iPsi_1}{\varphi}{} \bat{r_b} -\pder{\iPsi_2}{\varphi}{} \bat{r_b} \Bigg) = -\varepsilon \frac{m}{n^2 R} \Big( n^2 -1 \Big) \dot{f} (\varphi) \sin(m \varphi) \\ & \; \; \; \; \; \; \; \, +\varepsilon^2 \Bigg( \frac{2 m}{n^2 R} \Big( n^2 -1 \Big) f (\varphi) \dot{f} (\varphi) \sin(m \varphi) -\Big( n^2 -1 \Big) \sum_{q \neq m} \frac{q}{n^2 R} a_{1, q} \dot{f} (\varphi) \sin(q \varphi) -\sum_{q} \frac{q}{R} b_{1, q} \dot{f} (\varphi) \sin(q \varphi) \Bigg) \\ & \; \; \; \; \; \; \; \, +O \big( \varepsilon^3 \big),
		\end{split}
	\label{E_AngularDerivative_Expanded}
	\end{equation}
	\end{widetext}
	where the summation index has already been switched to $q$. Note that only the second boundary condition gets corrected by these terms and therefore merely $x_1$, $x_2$, $a_{1, p}$, and $a_{2, p}$ can receive corrections, $b_{1, p}$ and $b_{2, p}$ are extracted from the first boundary condition.

The last step in constructing the PT formulas is a Fourier analysis. By multiplying a boundary condition by $\cos(m \varphi)$ or $\cos(p \varphi)$ and integrating from $-\pi$ to $\pi$ the coupling integrals $C_{q p}$ and $D_{q p}$ are introduced and the frequencies and coefficients can be calculated. This is because each term occurs once in combination with $\cos(m \varphi)$ or in a sum over $q$ combined with $\cos(q \varphi)$. By Fourier analysis the $\varphi$-dependence is eliminated and the sums break down, leaving one of the desired values separately. Because all integrands are symmetric, the integration interval $0$ to $\pi$ can be used instead of $-\pi$ to $\pi$.

The corrections listed in Eqs.~\eqref{E_Eigenvalue_XOne} and \eqref{E_Eigenvalue_XTwo} are acquired from the first and second order of the expansion~\eqref{E_AngularDerivative_Expanded}, respectively. $x_1$ and $x_2$ follow by analyzing with $\cos(m \varphi)$, whereas analysis with $\cos(p \varphi)$, $p \neq m$, gives $a_{1, p}$ and $a_{2, p}$. The factor $R/(x_0 V_m)$, by which Eq.~\eqref{E_AngularDerivative_Expanded} has to be multiplied to get the correct contributions to Eqs.~\eqref{E_Eigenvalue_XOne} and \eqref{E_Eigenvalue_XTwo}, follow from the full equations for $x_1$ and $x_2$ that are not shown here. Note that an additional $x_0$ has to be factored out. The same applies for the correct contributions to Eqs.~\eqref{E_Coefficient_AOne} and \eqref{E_Coefficient_ATwo} with $a_{1, p}$, $a_{2, p}$, and the factor $R/(x_0 T_p)$. Furthermore, in the full formulas $x_1$ and $a_{1, p}$, as well as $x_2$ and $a_{2, p}$, have different signs, which is why those differ in the corrections too.

Now consider the difference for the parities. For the symmetric wave functions the first derivative of the cosine with respect to $\varphi$ produces a negative sign. This does not appear for the sines of the antisymmetric wave functions. This means that the sign of the corrections has to be switched in addition to the usual interchanging of the sines and cosines. With the signs $\pm$ and $\mp$ both parities can still be presented in one set of formulas.


\begin{thebibliography}{47}%
\makeatletter
\providecommand \@ifxundefined [1]{%
 \@ifx{#1\undefined}
}%
\providecommand \@ifnum [1]{%
 \ifnum #1\expandafter \@firstoftwo
 \else \expandafter \@secondoftwo
 \fi
}%
\providecommand \@ifx [1]{%
 \ifx #1\expandafter \@firstoftwo
 \else \expandafter \@secondoftwo
 \fi
}%
\providecommand \natexlab [1]{#1}%
\providecommand \enquote  [1]{``#1''}%
\providecommand \bibnamefont  [1]{#1}%
\providecommand \bibfnamefont [1]{#1}%
\providecommand \citenamefont [1]{#1}%
\providecommand \href@noop [0]{\@secondoftwo}%
\providecommand \href [0]{\begingroup \@sanitize@url \@href}%
\providecommand \@href[1]{\@@startlink{#1}\@@href}%
\providecommand \@@href[1]{\endgroup#1\@@endlink}%
\providecommand \@sanitize@url [0]{\catcode `\\12\catcode `\$12\catcode
  `\&12\catcode `\#12\catcode `\^12\catcode `\_12\catcode `\%12\relax}%
\providecommand \@@startlink[1]{}%
\providecommand \@@endlink[0]{}%
\providecommand \url  [0]{\begingroup\@sanitize@url \@url }%
\providecommand \@url [1]{\endgroup\@href {#1}{\urlprefix }}%
\providecommand \urlprefix  [0]{URL }%
\providecommand \Eprint [0]{\href }%
\providecommand \doibase [0]{http://dx.doi.org/}%
\providecommand \selectlanguage [0]{\@gobble}%
\providecommand \bibinfo  [0]{\@secondoftwo}%
\providecommand \bibfield  [0]{\@secondoftwo}%
\providecommand \translation [1]{[#1]}%
\providecommand \BibitemOpen [0]{}%
\providecommand \bibitemStop [0]{}%
\providecommand \bibitemNoStop [0]{.\EOS\space}%
\providecommand \EOS [0]{\spacefactor3000\relax}%
\providecommand \BibitemShut  [1]{\csname bibitem#1\endcsname}%
\let\auto@bib@innerbib\@empty
\bibitem [{\citenamefont {Vahala}(2003)}]{Vahala03}%
  \BibitemOpen
  \bibfield  {author} {\bibinfo {author} {\bibfnamefont {K.~J.}\ \bibnamefont
  {Vahala}},\ }\href {\doibase 10.1038/nature01939} {\bibfield  {journal}
  {\bibinfo  {journal} {Nature (London)}\ }\textbf {\bibinfo {volume} {424}},\
  \bibinfo {pages} {839} (\bibinfo {year} {2003})}\BibitemShut {NoStop}%
\bibitem [{\citenamefont {Zhu}\ \emph {et~al.}(2010)\citenamefont {Zhu},
  \citenamefont {{\"O}zdemir}, \citenamefont {Xiao}, \citenamefont {Li},
  \citenamefont {He}, \citenamefont {Chen},\ and\ \citenamefont
  {Yang}}]{ZOX10}%
  \BibitemOpen
  \bibfield  {author} {\bibinfo {author} {\bibfnamefont {J.}~\bibnamefont
  {Zhu}}, \bibinfo {author} {\bibfnamefont {{\c{S}}.~K.}\ \bibnamefont
  {{\"O}zdemir}}, \bibinfo {author} {\bibfnamefont {Y.-F.}\ \bibnamefont
  {Xiao}}, \bibinfo {author} {\bibfnamefont {L.}~\bibnamefont {Li}}, \bibinfo
  {author} {\bibfnamefont {L.}~\bibnamefont {He}}, \bibinfo {author}
  {\bibfnamefont {D.-R.}\ \bibnamefont {Chen}}, \ and\ \bibinfo {author}
  {\bibfnamefont {L.}~\bibnamefont {Yang}},\ }\href@noop {} {\bibfield
  {journal} {\bibinfo  {journal} {Nature Photon.}\ }\textbf {\bibinfo {volume}
  {4}},\ \bibinfo {pages} {46} (\bibinfo {year} {2010})}\BibitemShut {NoStop}%
\bibitem [{\citenamefont {Vollmer}\ and\ \citenamefont {Yang}(2012)}]{VY12}%
  \BibitemOpen
  \bibfield  {author} {\bibinfo {author} {\bibfnamefont {F.}~\bibnamefont
  {Vollmer}}\ and\ \bibinfo {author} {\bibfnamefont {L.}~\bibnamefont {Yang}},\
  }\href@noop {} {\bibfield  {journal} {\bibinfo  {journal} {Nanophotonics}\
  }\textbf {\bibinfo {volume} {1}},\ \bibinfo {pages} {267} (\bibinfo {year}
  {2012})}\BibitemShut {NoStop}%
\bibitem [{\citenamefont {{M. R. Foreman}}\ \emph {et~al.}(2015)\citenamefont
  {{M. R. Foreman}}, \citenamefont {{J. D. Swaim}},\ and\ \citenamefont {{F.
  Vollmer}}}]{FSV15}%
  \BibitemOpen
  \bibfield  {author} {\bibinfo {author} {\bibnamefont {{M. R. Foreman}}},
  \bibinfo {author} {\bibnamefont {{J. D. Swaim}}}, \ and\ \bibinfo {author}
  {\bibnamefont {{F. Vollmer}}},\ }\href {\doibase 10.1364/AOP.7.000168}
  {\bibfield  {journal} {\bibinfo  {journal} {Adv. Opt. Photon.}\ }\textbf
  {\bibinfo {volume} {7}},\ \bibinfo {pages} {168} (\bibinfo {year}
  {2015})}\BibitemShut {NoStop}%
\bibitem [{\citenamefont {Reitzenstein}\ \emph {et~al.}(2006)\citenamefont
  {Reitzenstein}, \citenamefont {Bazhenov}, \citenamefont {Gorbunov},
  \citenamefont {Hofmann}, \citenamefont {M{\"u}nch}, \citenamefont
  {L{\"o}ffler}, \citenamefont {Kamp}, \citenamefont {Reithmaier},
  \citenamefont {Kulakovskii},\ and\ \citenamefont {Forchel}}]{Reitzenstein06}%
  \BibitemOpen
  \bibfield  {author} {\bibinfo {author} {\bibfnamefont {S.}~\bibnamefont
  {Reitzenstein}}, \bibinfo {author} {\bibfnamefont {A.}~\bibnamefont
  {Bazhenov}}, \bibinfo {author} {\bibfnamefont {A.}~\bibnamefont {Gorbunov}},
  \bibinfo {author} {\bibfnamefont {C.}~\bibnamefont {Hofmann}}, \bibinfo
  {author} {\bibfnamefont {S.}~\bibnamefont {M{\"u}nch}}, \bibinfo {author}
  {\bibfnamefont {A.}~\bibnamefont {L{\"o}ffler}}, \bibinfo {author}
  {\bibfnamefont {M.}~\bibnamefont {Kamp}}, \bibinfo {author} {\bibfnamefont
  {J.~P.}\ \bibnamefont {Reithmaier}}, \bibinfo {author} {\bibfnamefont
  {V.~D.}\ \bibnamefont {Kulakovskii}}, \ and\ \bibinfo {author} {\bibfnamefont
  {A.}~\bibnamefont {Forchel}},\ }\href@noop {} {\bibfield  {journal} {\bibinfo
   {journal} {Appl. Phys. Lett.}\ }\textbf {\bibinfo {volume} {89}},\ \bibinfo
  {pages} {051107} (\bibinfo {year} {2006})}\BibitemShut {NoStop}%
\bibitem [{\citenamefont {Ulrich}\ \emph {et~al.}(2007)\citenamefont {Ulrich},
  \citenamefont {Gies}, \citenamefont {Ates}, \citenamefont {Wiersig},
  \citenamefont {Reitzenstein}, \citenamefont {Hofmann}, \citenamefont
  {L{\"o}ffler}, \citenamefont {Forchel}, \citenamefont {Jahnke},\ and\
  \citenamefont {Michler}}]{Ulrich06}%
  \BibitemOpen
  \bibfield  {author} {\bibinfo {author} {\bibfnamefont {S.~M.}\ \bibnamefont
  {Ulrich}}, \bibinfo {author} {\bibfnamefont {C.}~\bibnamefont {Gies}},
  \bibinfo {author} {\bibfnamefont {S.}~\bibnamefont {Ates}}, \bibinfo {author}
  {\bibfnamefont {J.}~\bibnamefont {Wiersig}}, \bibinfo {author} {\bibfnamefont
  {S.}~\bibnamefont {Reitzenstein}}, \bibinfo {author} {\bibfnamefont
  {C.}~\bibnamefont {Hofmann}}, \bibinfo {author} {\bibfnamefont
  {A.}~\bibnamefont {L{\"o}ffler}}, \bibinfo {author} {\bibfnamefont
  {A.}~\bibnamefont {Forchel}}, \bibinfo {author} {\bibfnamefont
  {F.}~\bibnamefont {Jahnke}}, \ and\ \bibinfo {author} {\bibfnamefont
  {P.}~\bibnamefont {Michler}},\ }\href@noop {} {\bibfield  {journal} {\bibinfo
   {journal} {Phys. Rev. Lett.}\ }\textbf {\bibinfo {volume} {98}},\ \bibinfo
  {pages} {043906} (\bibinfo {year} {2007})}\BibitemShut {NoStop}%
\bibitem [{\citenamefont {McCall}\ \emph {et~al.}(1992)\citenamefont {McCall},
  \citenamefont {Levi}, \citenamefont {Slusher}, \citenamefont {Pearton},\ and\
  \citenamefont {Logan}}]{MLSGPL92}%
  \BibitemOpen
  \bibfield  {author} {\bibinfo {author} {\bibfnamefont {S.~L.}\ \bibnamefont
  {McCall}}, \bibinfo {author} {\bibfnamefont {A.~F.~J.}\ \bibnamefont {Levi}},
  \bibinfo {author} {\bibfnamefont {R.~E.}\ \bibnamefont {Slusher}}, \bibinfo
  {author} {\bibfnamefont {S.~J.}\ \bibnamefont {Pearton}}, \ and\ \bibinfo
  {author} {\bibfnamefont {R.~A.}\ \bibnamefont {Logan}},\ }\href@noop {}
  {\bibfield  {journal} {\bibinfo  {journal} {Appl. Phys. Lett.}\ }\textbf
  {\bibinfo {volume} {60}},\ \bibinfo {pages} {289} (\bibinfo {year}
  {1992})}\BibitemShut {NoStop}%
\bibitem [{\citenamefont {Slusher}\ \emph {et~al.}(1993)\citenamefont
  {Slusher}, \citenamefont {Levi}, \citenamefont {Mohideen}, \citenamefont
  {McCall}, \citenamefont {Pearton},\ and\ \citenamefont {Logan}}]{SLMMOL93}%
  \BibitemOpen
  \bibfield  {author} {\bibinfo {author} {\bibfnamefont {R.~E.}\ \bibnamefont
  {Slusher}}, \bibinfo {author} {\bibfnamefont {A.~F.~J.}\ \bibnamefont
  {Levi}}, \bibinfo {author} {\bibfnamefont {U.}~\bibnamefont {Mohideen}},
  \bibinfo {author} {\bibfnamefont {S.~L.}\ \bibnamefont {McCall}}, \bibinfo
  {author} {\bibfnamefont {S.~J.}\ \bibnamefont {Pearton}}, \ and\ \bibinfo
  {author} {\bibfnamefont {R.~A.}\ \bibnamefont {Logan}},\ }\href@noop {}
  {\bibfield  {journal} {\bibinfo  {journal} {Appl. Phys. Lett.}\ }\textbf
  {\bibinfo {volume} {63}},\ \bibinfo {pages} {1310} (\bibinfo {year}
  {1993})}\BibitemShut {NoStop}%
\bibitem [{\citenamefont {Witzany}\ \emph {et~al.}(2011)\citenamefont
  {Witzany}, \citenamefont {Ro{\ss}bach}, \citenamefont {Schulz}, \citenamefont
  {Jetter}, \citenamefont {Michler}, \citenamefont {Liu}, \citenamefont {Hu},
  \citenamefont {Wiersig},\ and\ \citenamefont {Jahnke}}]{WRS11}%
  \BibitemOpen
  \bibfield  {author} {\bibinfo {author} {\bibfnamefont {M.}~\bibnamefont
  {Witzany}}, \bibinfo {author} {\bibfnamefont {R.}~\bibnamefont
  {Ro{\ss}bach}}, \bibinfo {author} {\bibfnamefont {W.-M.}\ \bibnamefont
  {Schulz}}, \bibinfo {author} {\bibfnamefont {M.}~\bibnamefont {Jetter}},
  \bibinfo {author} {\bibfnamefont {P.}~\bibnamefont {Michler}}, \bibinfo
  {author} {\bibfnamefont {T.-L.}\ \bibnamefont {Liu}}, \bibinfo {author}
  {\bibfnamefont {E.}~\bibnamefont {Hu}}, \bibinfo {author} {\bibfnamefont
  {J.}~\bibnamefont {Wiersig}}, \ and\ \bibinfo {author} {\bibfnamefont
  {F.}~\bibnamefont {Jahnke}},\ }\href@noop {} {\bibfield  {journal} {\bibinfo
  {journal} {Phys. Rev. B}\ }\textbf {\bibinfo {volume} {83}},\ \bibinfo
  {pages} {205305} (\bibinfo {year} {2011})}\BibitemShut {NoStop}%
\bibitem [{\citenamefont {N{\"o}ckel}\ and\ \citenamefont
  {Stone}(1997)}]{ND97}%
  \BibitemOpen
  \bibfield  {author} {\bibinfo {author} {\bibfnamefont {J.~U.}\ \bibnamefont
  {N{\"o}ckel}}\ and\ \bibinfo {author} {\bibfnamefont {A.~D.}\ \bibnamefont
  {Stone}},\ }\href {\doibase 10.1038/385045a0} {\bibfield  {journal} {\bibinfo
   {journal} {Nature (London)}\ }\textbf {\bibinfo {volume} {385}},\ \bibinfo
  {pages} {45} (\bibinfo {year} {1997})}\BibitemShut {NoStop}%
\bibitem [{\citenamefont {Gmachl}\ \emph {et~al.}(1998)\citenamefont {Gmachl},
  \citenamefont {Capasso}, \citenamefont {Narimanov}, \citenamefont
  {N{\"o}ckel}, \citenamefont {Stone}, \citenamefont {Faist}, \citenamefont
  {Sivco},\ and\ \citenamefont {Cho}}]{GCNNSFSC98}%
  \BibitemOpen
  \bibfield  {author} {\bibinfo {author} {\bibfnamefont {C.}~\bibnamefont
  {Gmachl}}, \bibinfo {author} {\bibfnamefont {F.}~\bibnamefont {Capasso}},
  \bibinfo {author} {\bibfnamefont {E.~E.}\ \bibnamefont {Narimanov}}, \bibinfo
  {author} {\bibfnamefont {J.~U.}\ \bibnamefont {N{\"o}ckel}}, \bibinfo
  {author} {\bibfnamefont {A.~D.}\ \bibnamefont {Stone}}, \bibinfo {author}
  {\bibfnamefont {J.}~\bibnamefont {Faist}}, \bibinfo {author} {\bibfnamefont
  {D.~L.}\ \bibnamefont {Sivco}}, \ and\ \bibinfo {author} {\bibfnamefont
  {A.~Y.}\ \bibnamefont {Cho}},\ }\href@noop {} {\bibfield  {journal} {\bibinfo
   {journal} {Science}\ }\textbf {\bibinfo {volume} {280}},\ \bibinfo {pages}
  {1556} (\bibinfo {year} {1998})}\BibitemShut {NoStop}%
\bibitem [{\citenamefont {Wiersig}\ and\ \citenamefont
  {Hentschel}(2008)}]{WH08}%
  \BibitemOpen
  \bibfield  {author} {\bibinfo {author} {\bibfnamefont {J.}~\bibnamefont
  {Wiersig}}\ and\ \bibinfo {author} {\bibfnamefont {M.}~\bibnamefont
  {Hentschel}},\ }\href {\doibase 10.1103/PhysRevLett.100.033901} {\bibfield
  {journal} {\bibinfo  {journal} {Phys. Rev. Lett.}\ }\textbf {\bibinfo
  {volume} {100}},\ \bibinfo {pages} {033901} (\bibinfo {year}
  {2008})}\BibitemShut {NoStop}%
\bibitem [{\citenamefont {Shinohara}\ \emph {et~al.}(2009)\citenamefont
  {Shinohara}, \citenamefont {Hentschel}, \citenamefont {Wiersig},
  \citenamefont {Sasaki},\ and\ \citenamefont {Harayama}}]{SHH09}%
  \BibitemOpen
  \bibfield  {author} {\bibinfo {author} {\bibfnamefont {S.}~\bibnamefont
  {Shinohara}}, \bibinfo {author} {\bibfnamefont {M.}~\bibnamefont
  {Hentschel}}, \bibinfo {author} {\bibfnamefont {J.}~\bibnamefont {Wiersig}},
  \bibinfo {author} {\bibfnamefont {T.}~\bibnamefont {Sasaki}}, \ and\ \bibinfo
  {author} {\bibfnamefont {T.}~\bibnamefont {Harayama}},\ }\href {\doibase
  10.1103/PhysRevA.80.031801} {\bibfield  {journal} {\bibinfo  {journal} {Phys.
  Rev. A}\ }\textbf {\bibinfo {volume} {80}},\ \bibinfo {pages} {031801(R)}
  (\bibinfo {year} {2009})}\BibitemShut {NoStop}%
\bibitem [{\citenamefont {Wang}\ \emph {et~al.}(2009)\citenamefont {Wang},
  \citenamefont {Yan}, \citenamefont {Diehl}, \citenamefont {Hentschel},
  \citenamefont {Wiersig}, \citenamefont {Yu}, \citenamefont {Pfl{\"u}gl},
  \citenamefont {Belkin}, \citenamefont {Edamura}, \citenamefont {Yamanishi},
  \citenamefont {Kan},\ and\ \citenamefont {Capasso}}]{WYD09}%
  \BibitemOpen
  \bibfield  {author} {\bibinfo {author} {\bibfnamefont {Q.~J.}\ \bibnamefont
  {Wang}}, \bibinfo {author} {\bibfnamefont {C.}~\bibnamefont {Yan}}, \bibinfo
  {author} {\bibfnamefont {L.}~\bibnamefont {Diehl}}, \bibinfo {author}
  {\bibfnamefont {M.}~\bibnamefont {Hentschel}}, \bibinfo {author}
  {\bibfnamefont {J.}~\bibnamefont {Wiersig}}, \bibinfo {author} {\bibfnamefont
  {N.}~\bibnamefont {Yu}}, \bibinfo {author} {\bibfnamefont {C.}~\bibnamefont
  {Pfl{\"u}gl}}, \bibinfo {author} {\bibfnamefont {M.~A.}\ \bibnamefont
  {Belkin}}, \bibinfo {author} {\bibfnamefont {T.}~\bibnamefont {Edamura}},
  \bibinfo {author} {\bibfnamefont {M.}~\bibnamefont {Yamanishi}}, \bibinfo
  {author} {\bibfnamefont {H.}~\bibnamefont {Kan}}, \ and\ \bibinfo {author}
  {\bibfnamefont {F.}~\bibnamefont {Capasso}},\ }\href@noop {} {\bibfield
  {journal} {\bibinfo  {journal} {New J. Phys.}\ }\textbf {\bibinfo {volume}
  {11}},\ \bibinfo {pages} {125018} (\bibinfo {year} {2009})}\BibitemShut
  {NoStop}%
\bibitem [{\citenamefont {Song}\ \emph {et~al.}(2009)\citenamefont {Song},
  \citenamefont {Fang}, \citenamefont {Liu}, \citenamefont {Ho}, \citenamefont
  {Solomon},\ and\ \citenamefont {Cao}}]{SCL09}%
  \BibitemOpen
  \bibfield  {author} {\bibinfo {author} {\bibfnamefont {Q.~H.}\ \bibnamefont
  {Song}}, \bibinfo {author} {\bibfnamefont {W.}~\bibnamefont {Fang}}, \bibinfo
  {author} {\bibfnamefont {B.}~\bibnamefont {Liu}}, \bibinfo {author}
  {\bibfnamefont {S.~T.}\ \bibnamefont {Ho}}, \bibinfo {author} {\bibfnamefont
  {G.~S.}\ \bibnamefont {Solomon}}, \ and\ \bibinfo {author} {\bibfnamefont
  {H.}~\bibnamefont {Cao}},\ }\href {\doibase 10.1103/PhysRevA.80.041807}
  {\bibfield  {journal} {\bibinfo  {journal} {Phys. Rev. A}\ }\textbf {\bibinfo
  {volume} {80}},\ \bibinfo {pages} {041807(R)} (\bibinfo {year}
  {2009})}\BibitemShut {NoStop}%
\bibitem [{\citenamefont {Yan}\ \emph {et~al.}(2009)\citenamefont {Yan},
  \citenamefont {Wang}, \citenamefont {Diehl}, \citenamefont {Hentschel},
  \citenamefont {Wiersig}, \citenamefont {Yu}, \citenamefont {Pf{\"u}gl},
  \citenamefont {Capasso}, \citenamefont {Belkin}, \citenamefont {Edamura},
  \citenamefont {Yamanishi},\ and\ \citenamefont {Kan}}]{YWD09}%
  \BibitemOpen
  \bibfield  {author} {\bibinfo {author} {\bibfnamefont {C.}~\bibnamefont
  {Yan}}, \bibinfo {author} {\bibfnamefont {Q.~J.}\ \bibnamefont {Wang}},
  \bibinfo {author} {\bibfnamefont {L.}~\bibnamefont {Diehl}}, \bibinfo
  {author} {\bibfnamefont {M.}~\bibnamefont {Hentschel}}, \bibinfo {author}
  {\bibfnamefont {J.}~\bibnamefont {Wiersig}}, \bibinfo {author} {\bibfnamefont
  {N.}~\bibnamefont {Yu}}, \bibinfo {author} {\bibnamefont {Pf{\"u}gl}},
  \bibinfo {author} {\bibfnamefont {F.}~\bibnamefont {Capasso}}, \bibinfo
  {author} {\bibfnamefont {M.~A.}\ \bibnamefont {Belkin}}, \bibinfo {author}
  {\bibfnamefont {T.}~\bibnamefont {Edamura}}, \bibinfo {author} {\bibfnamefont
  {M.}~\bibnamefont {Yamanishi}}, \ and\ \bibinfo {author} {\bibfnamefont
  {H.}~\bibnamefont {Kan}},\ }\href {\doibase 10.1063/1.3153276} {\bibfield
  {journal} {\bibinfo  {journal} {Appl. Phys. Lett.}\ }\textbf {\bibinfo
  {volume} {94}},\ \bibinfo {pages} {251101} (\bibinfo {year}
  {2009})}\BibitemShut {NoStop}%
\bibitem [{\citenamefont {Albert}\ \emph {et~al.}(2012)\citenamefont {Albert},
  \citenamefont {Hopfmann}, \citenamefont {Ebersp{\"a}cher}, \citenamefont
  {Arnold}, \citenamefont {Emmerling}, \citenamefont {Schneider}, \citenamefont
  {H{\"o}fling}, \citenamefont {Forchel}, \citenamefont {Kamp}, \citenamefont
  {Wiersig},\ and\ \citenamefont {Reitzenstein}}]{AHE12}%
  \BibitemOpen
  \bibfield  {author} {\bibinfo {author} {\bibfnamefont {F.}~\bibnamefont
  {Albert}}, \bibinfo {author} {\bibfnamefont {C.}~\bibnamefont {Hopfmann}},
  \bibinfo {author} {\bibfnamefont {A.}~\bibnamefont {Ebersp{\"a}cher}},
  \bibinfo {author} {\bibfnamefont {F.}~\bibnamefont {Arnold}}, \bibinfo
  {author} {\bibfnamefont {M.}~\bibnamefont {Emmerling}}, \bibinfo {author}
  {\bibfnamefont {C.}~\bibnamefont {Schneider}}, \bibinfo {author}
  {\bibfnamefont {S.}~\bibnamefont {H{\"o}fling}}, \bibinfo {author}
  {\bibfnamefont {A.}~\bibnamefont {Forchel}}, \bibinfo {author} {\bibfnamefont
  {M.}~\bibnamefont {Kamp}}, \bibinfo {author} {\bibfnamefont {J.}~\bibnamefont
  {Wiersig}}, \ and\ \bibinfo {author} {\bibfnamefont {S.}~\bibnamefont
  {Reitzenstein}},\ }\href {\doibase 10.1063/1.4733726} {\bibfield  {journal}
  {\bibinfo  {journal} {Appl. Phys. Lett.}\ }\textbf {\bibinfo {volume}
  {101}},\ \bibinfo {pages} {021116} (\bibinfo {year} {2012})}\BibitemShut
  {NoStop}%
\bibitem [{\citenamefont {{S. V. Boriskina}}\ \emph {et~al.}(2006)\citenamefont
  {{S. V. Boriskina}}, \citenamefont {{T. M. Benson}}, \citenamefont {{P. D.
  Sewell}},\ and\ \citenamefont {{A. I. Nosich}}}]{BBS06b}%
  \BibitemOpen
  \bibfield  {author} {\bibinfo {author} {\bibnamefont {{S. V. Boriskina}}},
  \bibinfo {author} {\bibnamefont {{T. M. Benson}}}, \bibinfo {author}
  {\bibnamefont {{P. D. Sewell}}}, \ and\ \bibinfo {author} {\bibnamefont {{A.
  I. Nosich}}},\ }\href {\doibase 10.1109/JSTQE.2006.882662} {\bibfield
  {journal} {\bibinfo  {journal} {IEEE J. Sel. Top. Quantum Electron.}\
  }\textbf {\bibinfo {volume} {12}},\ \bibinfo {pages} {1175} (\bibinfo {year}
  {2006})}\BibitemShut {NoStop}%
\bibitem [{\citenamefont {Wang}\ \emph {et~al.}(2010)\citenamefont {Wang},
  \citenamefont {Yan}, \citenamefont {Yu}, \citenamefont {Unterhinninghofen},
  \citenamefont {Wiersig}, \citenamefont {Pfl{\"u}gl}, \citenamefont {Diehl},
  \citenamefont {Edamura}, \citenamefont {Yamanishi}, \citenamefont {Kan},\
  and\ \citenamefont {Capasso}}]{WYY10}%
  \BibitemOpen
  \bibfield  {author} {\bibinfo {author} {\bibfnamefont {Q.~J.}\ \bibnamefont
  {Wang}}, \bibinfo {author} {\bibfnamefont {C.}~\bibnamefont {Yan}}, \bibinfo
  {author} {\bibfnamefont {N.}~\bibnamefont {Yu}}, \bibinfo {author}
  {\bibfnamefont {J.}~\bibnamefont {Unterhinninghofen}}, \bibinfo {author}
  {\bibfnamefont {J.}~\bibnamefont {Wiersig}}, \bibinfo {author} {\bibfnamefont
  {C.}~\bibnamefont {Pfl{\"u}gl}}, \bibinfo {author} {\bibfnamefont
  {L.}~\bibnamefont {Diehl}}, \bibinfo {author} {\bibfnamefont
  {T.}~\bibnamefont {Edamura}}, \bibinfo {author} {\bibfnamefont
  {M.}~\bibnamefont {Yamanishi}}, \bibinfo {author} {\bibfnamefont
  {H.}~\bibnamefont {Kan}}, \ and\ \bibinfo {author} {\bibfnamefont
  {F.}~\bibnamefont {Capasso}},\ }\href@noop {} {\bibfield  {journal} {\bibinfo
   {journal} {Proc. Natl. Acad. Sci. USA}\ }\textbf {\bibinfo {volume} {107}},\
  \bibinfo {pages} {22407} (\bibinfo {year} {2010})}\BibitemShut {NoStop}%
\bibitem [{\citenamefont {Schermer}\ \emph {et~al.}(2015)\citenamefont
  {Schermer}, \citenamefont {Bittner}, \citenamefont {Singh}, \citenamefont
  {Ulysee}, \citenamefont {Lebental},\ and\ \citenamefont {Wiersig}}]{SBS15}%
  \BibitemOpen
  \bibfield  {author} {\bibinfo {author} {\bibfnamefont {M.}~\bibnamefont
  {Schermer}}, \bibinfo {author} {\bibfnamefont {S.}~\bibnamefont {Bittner}},
  \bibinfo {author} {\bibfnamefont {G.}~\bibnamefont {Singh}}, \bibinfo
  {author} {\bibfnamefont {C.}~\bibnamefont {Ulysee}}, \bibinfo {author}
  {\bibfnamefont {M.}~\bibnamefont {Lebental}}, \ and\ \bibinfo {author}
  {\bibfnamefont {J.}~\bibnamefont {Wiersig}},\ }\href@noop {} {\bibfield
  {journal} {\bibinfo  {journal} {Appl. Phys. Lett.}\ }\textbf {\bibinfo
  {volume} {106}},\ \bibinfo {pages} {101107} (\bibinfo {year}
  {2015})}\BibitemShut {NoStop}%
\bibitem [{\citenamefont {Jiang}\ \emph {et~al.}(2017)\citenamefont {Jiang},
  \citenamefont {Shao}, \citenamefont {Zhang}, \citenamefont {Yi},
  \citenamefont {Wiersig}, \citenamefont {Wang}, \citenamefont {Gong},
  \citenamefont {Loncar}, \citenamefont {Yang},\ and\ \citenamefont
  {Xiao}}]{JSZ17}%
  \BibitemOpen
  \bibfield  {author} {\bibinfo {author} {\bibfnamefont {X.}~\bibnamefont
  {Jiang}}, \bibinfo {author} {\bibfnamefont {L.}~\bibnamefont {Shao}},
  \bibinfo {author} {\bibfnamefont {S.-X.}\ \bibnamefont {Zhang}}, \bibinfo
  {author} {\bibfnamefont {X.}~\bibnamefont {Yi}}, \bibinfo {author}
  {\bibfnamefont {J.}~\bibnamefont {Wiersig}}, \bibinfo {author} {\bibfnamefont
  {L.}~\bibnamefont {Wang}}, \bibinfo {author} {\bibfnamefont {Q.}~\bibnamefont
  {Gong}}, \bibinfo {author} {\bibfnamefont {M.}~\bibnamefont {Loncar}},
  \bibinfo {author} {\bibfnamefont {L.}~\bibnamefont {Yang}}, \ and\ \bibinfo
  {author} {\bibfnamefont {Y.-F.}\ \bibnamefont {Xiao}},\ }\href@noop {}
  {\bibfield  {journal} {\bibinfo  {journal} {Science}\ }\textbf {\bibinfo
  {volume} {358}},\ \bibinfo {pages} {344} (\bibinfo {year}
  {2017})}\BibitemShut {NoStop}%
\bibitem [{\citenamefont {N{\"o}ckel}\ \emph {et~al.}(1994)\citenamefont
  {N{\"o}ckel}, \citenamefont {Stone},\ and\ \citenamefont
  {Chang}}]{Noeckel94}%
  \BibitemOpen
  \bibfield  {author} {\bibinfo {author} {\bibfnamefont {J.~U.}\ \bibnamefont
  {N{\"o}ckel}}, \bibinfo {author} {\bibfnamefont {A.~D.}\ \bibnamefont
  {Stone}}, \ and\ \bibinfo {author} {\bibfnamefont {R.~K.}\ \bibnamefont
  {Chang}},\ }\href@noop {} {\bibfield  {journal} {\bibinfo  {journal} {Opt.
  Lett.}\ }\textbf {\bibinfo {volume} {19}},\ \bibinfo {pages} {1693} (\bibinfo
  {year} {1994})}\BibitemShut {NoStop}%
\bibitem [{\citenamefont {Fujita}\ and\ \citenamefont {Baba}(2001)}]{FB01}%
  \BibitemOpen
  \bibfield  {author} {\bibinfo {author} {\bibfnamefont {M.}~\bibnamefont
  {Fujita}}\ and\ \bibinfo {author} {\bibfnamefont {T.}~\bibnamefont {Baba}},\
  }\href@noop {} {\bibfield  {journal} {\bibinfo  {journal} {IEEE J. Quantum
  Electron.}\ }\textbf {\bibinfo {volume} {37}},\ \bibinfo {pages} {1253}
  (\bibinfo {year} {2001})}\BibitemShut {NoStop}%
\bibitem [{\citenamefont {Fujita}\ and\ \citenamefont {Baba}(2002)}]{FB02}%
  \BibitemOpen
  \bibfield  {author} {\bibinfo {author} {\bibfnamefont {M.}~\bibnamefont
  {Fujita}}\ and\ \bibinfo {author} {\bibfnamefont {T.}~\bibnamefont {Baba}},\
  }\href@noop {} {\bibfield  {journal} {\bibinfo  {journal} {Appl. Phys.
  Lett.}\ }\textbf {\bibinfo {volume} {80}},\ \bibinfo {pages} {2051} (\bibinfo
  {year} {2002})}\BibitemShut {NoStop}%
\bibitem [{\citenamefont {Schlehahn}\ \emph {et~al.}(2013)\citenamefont
  {Schlehahn}, \citenamefont {Albert}, \citenamefont {Schneider}, \citenamefont
  {H{\"o}fling}, \citenamefont {Reitzenstein}, \citenamefont {Wiersig},\ and\
  \citenamefont {Kamp}}]{SAS13}%
  \BibitemOpen
  \bibfield  {author} {\bibinfo {author} {\bibfnamefont {A.}~\bibnamefont
  {Schlehahn}}, \bibinfo {author} {\bibfnamefont {F.}~\bibnamefont {Albert}},
  \bibinfo {author} {\bibfnamefont {C.}~\bibnamefont {Schneider}}, \bibinfo
  {author} {\bibfnamefont {S.}~\bibnamefont {H{\"o}fling}}, \bibinfo {author}
  {\bibfnamefont {S.}~\bibnamefont {Reitzenstein}}, \bibinfo {author}
  {\bibfnamefont {J.}~\bibnamefont {Wiersig}}, \ and\ \bibinfo {author}
  {\bibfnamefont {M.}~\bibnamefont {Kamp}},\ }\href@noop {} {\bibfield
  {journal} {\bibinfo  {journal} {Opt. Express}\ }\textbf {\bibinfo {volume}
  {21}},\ \bibinfo {pages} {15951} (\bibinfo {year} {2013})}\BibitemShut
  {NoStop}%
\bibitem [{\citenamefont {Cao}\ and\ \citenamefont {Wiersig}(2015)}]{CW15}%
  \BibitemOpen
  \bibfield  {author} {\bibinfo {author} {\bibfnamefont {H.}~\bibnamefont
  {Cao}}\ and\ \bibinfo {author} {\bibfnamefont {J.}~\bibnamefont {Wiersig}},\
  }\href {\doibase 10.1103/RevModPhys.87.61} {\bibfield  {journal} {\bibinfo
  {journal} {Rev. Mod. Phys.}\ }\textbf {\bibinfo {volume} {87}},\ \bibinfo
  {pages} {61} (\bibinfo {year} {2015})}\BibitemShut {NoStop}%
\bibitem [{\citenamefont {Dubertrand}\ \emph {et~al.}(2008)\citenamefont
  {Dubertrand}, \citenamefont {Bogomolny}, \citenamefont {Djellali},
  \citenamefont {Lebental},\ and\ \citenamefont {Schmit}}]{DBD08}%
  \BibitemOpen
  \bibfield  {author} {\bibinfo {author} {\bibfnamefont {R.}~\bibnamefont
  {Dubertrand}}, \bibinfo {author} {\bibfnamefont {E.}~\bibnamefont
  {Bogomolny}}, \bibinfo {author} {\bibfnamefont {N.}~\bibnamefont {Djellali}},
  \bibinfo {author} {\bibfnamefont {M.}~\bibnamefont {Lebental}}, \ and\
  \bibinfo {author} {\bibfnamefont {C.}~\bibnamefont {Schmit}},\ }\href
  {\doibase 10.1103/PhysRevA.77.013804} {\bibfield  {journal} {\bibinfo
  {journal} {Phys. Rev. A}\ }\textbf {\bibinfo {volume} {77}},\ \bibinfo
  {pages} {013804} (\bibinfo {year} {2008})}\BibitemShut {NoStop}%
\bibitem [{\citenamefont {Wiersig}(2012)}]{Wiersig12}%
  \BibitemOpen
  \bibfield  {author} {\bibinfo {author} {\bibfnamefont {J.}~\bibnamefont
  {Wiersig}},\ }\href {\doibase 10.1103/PhysRevA.85.063838} {\bibfield
  {journal} {\bibinfo  {journal} {Phys. Rev. A}\ }\textbf {\bibinfo {volume}
  {85}},\ \bibinfo {pages} {063838} (\bibinfo {year} {2012})}\BibitemShut
  {NoStop}%
\bibitem [{\citenamefont {Kraft}\ and\ \citenamefont {Wiersig}(2014)}]{KW14}%
  \BibitemOpen
  \bibfield  {author} {\bibinfo {author} {\bibfnamefont {M.}~\bibnamefont
  {Kraft}}\ and\ \bibinfo {author} {\bibfnamefont {J.}~\bibnamefont
  {Wiersig}},\ }\href {\doibase 10.1103/PhysRevA.89.023819} {\bibfield
  {journal} {\bibinfo  {journal} {Phys. Rev. A}\ }\textbf {\bibinfo {volume}
  {89}},\ \bibinfo {pages} {023819} (\bibinfo {year} {2014})}\BibitemShut
  {NoStop}%
\bibitem [{\citenamefont {Badel}(2015)}]{Badel15}%
  \BibitemOpen
  \bibfield  {author} {\bibinfo {author} {\bibfnamefont {M.}~\bibnamefont
  {Badel}},\ }\emph {\bibinfo {title} {St\"{o}rungstheoretische Analyse der
  Frequenzen optischer Moden in elliptisch verformten Mikroresonatoren}},\
  \href@noop {} {\bibinfo {type} {Bachelor {T}hesis}},\ \bibinfo  {school}
  {Otto-von-Guericke-Universit\"{a}t Magdeburg} (\bibinfo {year}
  {2015})\BibitemShut {NoStop}%
\bibitem [{\citenamefont {Kraft}\ and\ \citenamefont
  {Wiersig}(2016)}]{KraftW16}%
  \BibitemOpen
  \bibfield  {author} {\bibinfo {author} {\bibfnamefont {M.}~\bibnamefont
  {Kraft}}\ and\ \bibinfo {author} {\bibfnamefont {J.}~\bibnamefont
  {Wiersig}},\ }\href {\doibase 10.1103/PhysRevA.94.013851} {\bibfield
  {journal} {\bibinfo  {journal} {Phys. Rev. A}\ }\textbf {\bibinfo {volume}
  {94}},\ \bibinfo {pages} {013851} (\bibinfo {year} {2016})}\BibitemShut
  {NoStop}%
\bibitem [{\citenamefont {Kullig}\ and\ \citenamefont {Wiersig}(2016)}]{KW16c}%
  \BibitemOpen
  \bibfield  {author} {\bibinfo {author} {\bibfnamefont {J.}~\bibnamefont
  {Kullig}}\ and\ \bibinfo {author} {\bibfnamefont {J.}~\bibnamefont
  {Wiersig}},\ }\href {\doibase 10.1103/PhysRevA.94.043850} {\bibfield
  {journal} {\bibinfo  {journal} {Phys. Rev. A}\ }\textbf {\bibinfo {volume}
  {94}},\ \bibinfo {pages} {043850} (\bibinfo {year} {2016})}\BibitemShut
  {NoStop}%
\bibitem [{\citenamefont {Wiersig}\ and\ \citenamefont {Kullig}(2017)}]{WK17}%
  \BibitemOpen
  \bibfield  {author} {\bibinfo {author} {\bibfnamefont {J.}~\bibnamefont
  {Wiersig}}\ and\ \bibinfo {author} {\bibfnamefont {J.}~\bibnamefont
  {Kullig}},\ }\href {\doibase 10.1103/PhysRevA.95.053815} {\bibfield
  {journal} {\bibinfo  {journal} {Phys. Rev. A}\ }\textbf {\bibinfo {volume}
  {95}},\ \bibinfo {pages} {053815} (\bibinfo {year} {2017})}\BibitemShut
  {NoStop}%
\bibitem [{\citenamefont {Badel}(2017)}]{Badel17}%
  \BibitemOpen
  \bibfield  {author} {\bibinfo {author} {\bibfnamefont {M.}~\bibnamefont
  {Badel}},\ }\emph {\bibinfo {title} {Analyse und Vergleich der Tauglichkeit
  der TM- und TE-St\"{o}rungstheorien f\"{u}r Mikrodisks mit verschiedenen
  Verformungen}},\ \href@noop {} {\bibinfo {type} {Master {T}hesis}},\ \bibinfo
   {school} {Otto-von-Guericke-Universit\"{a}t Magdeburg} (\bibinfo {year}
  {2017})\BibitemShut {NoStop}%
\bibitem [{\citenamefont {Kullig}\ \emph {et~al.}(2018)\citenamefont {Kullig},
  \citenamefont {Yi},\ and\ \citenamefont {Wiersig}}]{KYW18}%
  \BibitemOpen
  \bibfield  {author} {\bibinfo {author} {\bibfnamefont {J.}~\bibnamefont
  {Kullig}}, \bibinfo {author} {\bibfnamefont {C.-H.}\ \bibnamefont {Yi}}, \
  and\ \bibinfo {author} {\bibfnamefont {J.}~\bibnamefont {Wiersig}},\
  }\href@noop {} {\bibfield  {journal} {\bibinfo  {journal} {Phys. Rev. A}\
  }\textbf {\bibinfo {volume} {98}},\ \bibinfo {pages} {023851} (\bibinfo
  {year} {2018})}\BibitemShut {NoStop}%
\bibitem [{\citenamefont {Ge}\ \emph {et~al.}(2013)\citenamefont {Ge},
  \citenamefont {Song}, \citenamefont {Redding},\ and\ \citenamefont
  {Cao}}]{GSR13}%
  \BibitemOpen
  \bibfield  {author} {\bibinfo {author} {\bibfnamefont {L.}~\bibnamefont
  {Ge}}, \bibinfo {author} {\bibfnamefont {Q.~H.}\ \bibnamefont {Song}},
  \bibinfo {author} {\bibfnamefont {B.}~\bibnamefont {Redding}}, \ and\
  \bibinfo {author} {\bibfnamefont {H.}~\bibnamefont {Cao}},\ }\href {\doibase
  10.1103/PhysRevA.87.023833} {\bibfield  {journal} {\bibinfo  {journal} {Phys.
  Rev. A}\ }\textbf {\bibinfo {volume} {87}},\ \bibinfo {pages} {023833}
  (\bibinfo {year} {2013})}\BibitemShut {NoStop}%
\bibitem [{\citenamefont {{S. V. Boriskina}}\ \emph {et~al.}(2003)\citenamefont
  {{S. V. Boriskina}}, \citenamefont {{T. M. Benson}}, \citenamefont {{P. D.
  Sewell}},\ and\ \citenamefont {{A. I. Nosich}}}]{BBS03}%
  \BibitemOpen
  \bibfield  {author} {\bibinfo {author} {\bibnamefont {{S. V. Boriskina}}},
  \bibinfo {author} {\bibnamefont {{T. M. Benson}}}, \bibinfo {author}
  {\bibnamefont {{P. D. Sewell}}}, \ and\ \bibinfo {author} {\bibnamefont {{A.
  I. Nosich}}},\ }\href {\doibase 10.1023/A:1022982024831} {\bibfield
  {journal} {\bibinfo  {journal} {Opt. Quant. Electron.}\ }\textbf {\bibinfo
  {volume} {35}},\ \bibinfo {pages} {545} (\bibinfo {year} {2003})}\BibitemShut
  {NoStop}%
\bibitem [{\citenamefont {Smotrova}\ \emph {et~al.}(2005)\citenamefont
  {Smotrova}, \citenamefont {Nosich}, \citenamefont {Benson},\ and\
  \citenamefont {Sewell}}]{SNB05}%
  \BibitemOpen
  \bibfield  {author} {\bibinfo {author} {\bibfnamefont {E.~I.}\ \bibnamefont
  {Smotrova}}, \bibinfo {author} {\bibfnamefont {A.~I.}\ \bibnamefont
  {Nosich}}, \bibinfo {author} {\bibfnamefont {T.~M.}\ \bibnamefont {Benson}},
  \ and\ \bibinfo {author} {\bibfnamefont {P.}~\bibnamefont {Sewell}},\ }\href
  {\doibase 10.1109/JSTQE.2005.853848} {\bibfield  {journal} {\bibinfo
  {journal} {IEEE J. Sel. Top. Quantum Electron.}\ }\textbf {\bibinfo {volume}
  {11}},\ \bibinfo {pages} {1135} (\bibinfo {year} {2005})}\BibitemShut
  {NoStop}%
\bibitem [{\citenamefont {Lebental}\ \emph {et~al.}(2007)\citenamefont
  {Lebental}, \citenamefont {Djellali}, \citenamefont {Arnaud}, \citenamefont
  {Lauret}, \citenamefont {Zyss}, \citenamefont {Dubertrand}, \citenamefont
  {Schmit},\ and\ \citenamefont {Bogomolny}}]{LDA07}%
  \BibitemOpen
  \bibfield  {author} {\bibinfo {author} {\bibfnamefont {M.}~\bibnamefont
  {Lebental}}, \bibinfo {author} {\bibfnamefont {N.}~\bibnamefont {Djellali}},
  \bibinfo {author} {\bibfnamefont {C.}~\bibnamefont {Arnaud}}, \bibinfo
  {author} {\bibfnamefont {J.-S.}\ \bibnamefont {Lauret}}, \bibinfo {author}
  {\bibfnamefont {J.}~\bibnamefont {Zyss}}, \bibinfo {author} {\bibfnamefont
  {R.}~\bibnamefont {Dubertrand}}, \bibinfo {author} {\bibfnamefont
  {C.}~\bibnamefont {Schmit}}, \ and\ \bibinfo {author} {\bibfnamefont
  {E.}~\bibnamefont {Bogomolny}},\ }\href {\doibase 10.1103/PhysRevA.76.023830}
  {\bibfield  {journal} {\bibinfo  {journal} {Phys. Rev. A}\ }\textbf {\bibinfo
  {volume} {76}},\ \bibinfo {pages} {023830} (\bibinfo {year}
  {2007})}\BibitemShut {NoStop}%
\bibitem [{\citenamefont {Bittner}\ \emph {et~al.}(2009)\citenamefont
  {Bittner}, \citenamefont {Dietz}, \citenamefont {Miski-Oglu}, \citenamefont
  {Iriarte}, \citenamefont {Richter},\ and\ \citenamefont
  {Sch{\"a}fer}}]{BDM09}%
  \BibitemOpen
  \bibfield  {author} {\bibinfo {author} {\bibfnamefont {S.}~\bibnamefont
  {Bittner}}, \bibinfo {author} {\bibfnamefont {B.}~\bibnamefont {Dietz}},
  \bibinfo {author} {\bibfnamefont {M.}~\bibnamefont {Miski-Oglu}}, \bibinfo
  {author} {\bibfnamefont {P.~O.}\ \bibnamefont {Iriarte}}, \bibinfo {author}
  {\bibfnamefont {A.}~\bibnamefont {Richter}}, \ and\ \bibinfo {author}
  {\bibfnamefont {F.}~\bibnamefont {Sch{\"a}fer}},\ }\href@noop {} {\bibfield
  {journal} {\bibinfo  {journal} {Phys. Rev. A}\ }\textbf {\bibinfo {volume}
  {80}},\ \bibinfo {pages} {023825} (\bibinfo {year} {2009})}\BibitemShut
  {NoStop}%
\bibitem [{\citenamefont {Lai}\ \emph {et~al.}(1990)\citenamefont {Lai},
  \citenamefont {Leung}, \citenamefont {Young}, \citenamefont {Barber},\ and\
  \citenamefont {Hill}}]{LLY90}%
  \BibitemOpen
  \bibfield  {author} {\bibinfo {author} {\bibfnamefont {H.~M.}\ \bibnamefont
  {Lai}}, \bibinfo {author} {\bibfnamefont {P.~T.}\ \bibnamefont {Leung}},
  \bibinfo {author} {\bibfnamefont {K.}~\bibnamefont {Young}}, \bibinfo
  {author} {\bibfnamefont {P.~W.}\ \bibnamefont {Barber}}, \ and\ \bibinfo
  {author} {\bibfnamefont {S.~C.}\ \bibnamefont {Hill}},\ }\href@noop {}
  {\bibfield  {journal} {\bibinfo  {journal} {Phys. Rev. A}\ }\textbf {\bibinfo
  {volume} {41}},\ \bibinfo {pages} {5187} (\bibinfo {year}
  {1990})}\BibitemShut {NoStop}%
\bibitem [{\citenamefont {van~den Berg}\ and\ \citenamefont
  {Fokkema}(1979)}]{BergFokkema79}%
  \BibitemOpen
  \bibfield  {author} {\bibinfo {author} {\bibfnamefont {P.~M.}\ \bibnamefont
  {van~den Berg}}\ and\ \bibinfo {author} {\bibfnamefont {J.~T.}\ \bibnamefont
  {Fokkema}},\ }\href {\doibase 10.1109/TAP.1979.1142152} {\bibfield  {journal}
  {\bibinfo  {journal} {IEEE Trans. Antennas Propag.}\ }\textbf {\bibinfo
  {volume} {27}},\ \bibinfo {pages} {577} (\bibinfo {year} {1979})}\BibitemShut
  {NoStop}%
\bibitem [{\citenamefont {Landau}(2008)}]{Landau08}%
  \BibitemOpen
  \bibfield  {author} {\bibinfo {author} {\bibfnamefont {R.~H.}\ \bibnamefont
  {Landau}},\ }\href@noop {} {\emph {\bibinfo {title} {Quantum Mechanics II}}}\
  (\bibinfo  {publisher} {John Wiley {\&} Sons},\ \bibinfo {address} {New
  York},\ \bibinfo {year} {2008})\BibitemShut {NoStop}%
\bibitem [{\citenamefont {Tureci}\ \emph {et~al.}(2005)\citenamefont {Tureci},
  \citenamefont {Schwefel}, \citenamefont {Jacqoud},\ and\ \citenamefont
  {Stone}}]{TSSJ05}%
  \BibitemOpen
  \bibfield  {author} {\bibinfo {author} {\bibfnamefont {H.~E.}\ \bibnamefont
  {Tureci}}, \bibinfo {author} {\bibfnamefont {H.~G.~L.}\ \bibnamefont
  {Schwefel}}, \bibinfo {author} {\bibfnamefont {P.}~\bibnamefont {Jacqoud}}, \
  and\ \bibinfo {author} {\bibfnamefont {A.~D.}\ \bibnamefont {Stone}},\ }\href
  {\doibase 10.1016/S0079-6638(05)47002-X} {\bibfield  {journal} {\bibinfo
  {journal} {Prog. Opt.}\ }\textbf {\bibinfo {volume} {47}},\ \bibinfo {pages}
  {75} (\bibinfo {year} {2005})}\BibitemShut {NoStop}%
\bibitem [{\citenamefont {Bogomolny}\ and\ \citenamefont
  {Dubertrand}(2012)}]{BD12}%
  \BibitemOpen
  \bibfield  {author} {\bibinfo {author} {\bibfnamefont {E.}~\bibnamefont
  {Bogomolny}}\ and\ \bibinfo {author} {\bibfnamefont {R.}~\bibnamefont
  {Dubertrand}},\ }\href@noop {} {\bibfield  {journal} {\bibinfo  {journal}
  {Phys. Rev. E}\ }\textbf {\bibinfo {volume} {86}},\ \bibinfo {pages} {026202}
  (\bibinfo {year} {2012})}\BibitemShut {NoStop}%
\bibitem [{\citenamefont {Wiersig}(2003)}]{Wiersig02b}%
  \BibitemOpen
  \bibfield  {author} {\bibinfo {author} {\bibfnamefont {J.}~\bibnamefont
  {Wiersig}},\ }\href@noop {} {\bibfield  {journal} {\bibinfo  {journal} {J.
  Opt. A: Pure Appl. Opt.}\ }\textbf {\bibinfo {volume} {5}},\ \bibinfo {pages}
  {53} (\bibinfo {year} {2003})}\BibitemShut {NoStop}%
\bibitem [{\citenamefont {Huy}\ \emph {et~al.}(2005)\citenamefont {Huy},
  \citenamefont {Morand},\ and\ \citenamefont {Benech}}]{HMB05}%
  \BibitemOpen
  \bibfield  {author} {\bibinfo {author} {\bibfnamefont {K.~P.}\ \bibnamefont
  {Huy}}, \bibinfo {author} {\bibfnamefont {A.}~\bibnamefont {Morand}}, \ and\
  \bibinfo {author} {\bibfnamefont {P.}~\bibnamefont {Benech}},\ }\href
  {\doibase 10.1109/JQE.2004.841498} {\bibfield  {journal} {\bibinfo  {journal}
  {IEEE J. Quantum Electron.}\ }\textbf {\bibinfo {volume} {41}},\ \bibinfo
  {pages} {357} (\bibinfo {year} {2005})}\BibitemShut {NoStop}%
\end{thebibliography}
%

\end{document}